\documentclass{emulateapj}

\newcommand{\beq}	{\begin{equation}}
\newcommand{\eeq}	{\end{equation}}
\newcommand{\beqa}	{\begin{eqnarray}}
\newcommand{\eeqa}	{\end{eqnarray}}
\newcommand{\calc}	{{\cal C}}
\newcommand{\calm}	{{\cal M}}

\newcommand{\caln}      {{\cal N}}
\newcommand{\calr}	{{\cal R}}

\newcommand{\avg}[1]    {{\langle #1 \rangle}} 
\newcommand{\e}	        {$^{-1}$}
\newcommand{\ee}	{$^{-2}$}
\newcommand{\eee}	{$^{-3}$}

\newcommand{\alfven}    {{Alfv$\acute{\rm e}$n }}

\newcommand{\alfvenic}  {{Alfv$\acute{\rm e}$nic }}
\newcommand{\alfvenicstop}  {{Alfv$\acute{\rm e}$nic}}

\newcommand{\chio}	{\chi_{i0}}

\newcommand{\gad}       {{\gamma_{\rm AD}}}

\newcommand{\htwo}	{H$_2$}
\newcommand{\mai}	{{\calm_{{\rm A}i}}}

\newcommand{\maic}      {{\calm_{{\rm A}i,c}^2}}
\newcommand{\muh}	{\mu_{\rm H}}
\newcommand{\nh}	{n_{\rm H}}
\newcommand{\rad}	{R_{\rm AD}}

\newcommand{\radc}	{R_{\rm AD,\,c}}

\newcommand{\tad}	{t_{\rm AD}}

\newcommand{\tf}	{t_f}

\newcommand{\vad}       {v_{\rm AD}}

\newcommand{\va}	{v_{\rm A}}

\newcommand{\avir}      {\alpha_{\rm vir}}

\newcommand\brms        {B_{\rm rms}}
\newcommand{\chiis}     {{\chi_i^*}}
\newcommand{\kphys}     {k_{\rm phys}}
\newcommand{\kinmax}     {k_{\rm in,\,max}}

\newcommand{\lad}        {{\ell_{\rm AD}}}

\newcommand{\lj}	{\lambda_{\rm J}}
\newcommand{\lpc}	{{\ell_{0,\,\rm pc}}}
\newcommand{\ma}	{{\calm_{\rm A}}}

\newcommand{\mug}	{\mu{\rm G}}
\newcommand{\nbh}	{\bar n_{\rm H}}
\newcommand{\nbht}	{\bar n_{\rm H,\, 3}}
\newcommand{\radl}	{R_{\rm AD}(\ell)}
\newcommand{\radlo}	{R_{\rm AD}(\ell_0)}
\newcommand{\rms}       {{\rm rms}}
\newcommand{\snt}       {\sigma_{\rm nt}}
\newcommand{\spc}	{\sigma_{\rm pc}}
\newcommand{\spcs}      {{\sigma_{\rm pc}^*}}

\newcommand{\clmf}  {ClMF}
\newcommand{\clmfs}  {ClMFs}
\newcommand{\gfit}	{\Gamma_{\rm fit}}
\newcommand{\ljt}	{\lambda_{\rm J,\, turb}}
\newcommand{\ljc}	{\lambda_{\rm J,\, core}}
\newcommand{\lws}	{{\rm LWS}}
\newcommand{\mbe}	{M_{\rm BE}}
\newcommand{\mj}	{M_{\rm J}}
\newcommand{\mjc}	{M_{\rm J,\, core}}
\newcommand{\mjt}	{M_{\rm J,\, turb}}

\newcommand{\nj}	{n_{\rm J}}
\newcommand{\njt}	{n_{\rm J,\, turb}}

\newcommand{\Nh}	{N_{\rm H}}
\newcommand{\Nhtt}	{N_{\rm H,22}}
\newcommand{\pc}		{{\rm pc}}

\newcommand{\turb}	{{\rm turb}}
\newcommand{\vffo}{v_{\rm ff}}

\newcommand{\lcode}	{\ell_{\rm code}}
\newcommand{\mcode}	{M_{\rm code}}

\newcommand{\tcode}	{t_{\rm code}}
\newcommand{\cstil}	{\tilde c_s}

\newcommand{\ltil}	{\tilde\ell}
\newcommand{\mtil}	{\tilde M}
\newcommand{\ttil}	{\tilde t}
\newcommand{\xeeq}	{x_{e,\,\rm eq}}
\newcommand{\zcr}	{\zeta_{\rm CR}}
\newcommand{\rhocore} {\rho_{\rm core}}
\newcommand{\mfr}       {$\mu_{\Phi,c}/\mu_{\Phi,0}$}

\newcommand{\rcz}	{R_{cz}}

\shorttitle{MHD Turbulence Simulations with Ambipolar Diffusion}
\shortauthors{McKee, Li, \& Klein}
\begin{document}

\title{Sub-\alfvenic Non-Ideal MHD Turbulence Simulations with Ambipolar Diffusion: II. 
Comparison with Observation, 
Clump Properties, and Scaling to Physical Units}
\author{Christopher F. McKee}
\affil{Physics Department and Astronomy Department, University of California,
    Berkeley, CA 94720; and Laboratoire d'Etudes du Rayonnement et de la Mati\`ere en Astrophysique, LERMA-LRA, Ecole Normale Superieure, 24 rue Lhomond, 75005 Paris, 
France}
\email{cmckee@astro.berkeley.edu}
\author{Pak Shing Li} \affil{Astronomy Department, University of California,
    Berkeley, CA 94720}
\email{psli@astron.berkeley.edu}
\and
\author{Richard I. Klein}
\affil{Astronomy Department, University of California,
    Berkeley, CA 94720; and Lawrence Livermore National Laboratory,\\
    P.O.Box 808, L-23, Livermore, CA 94550}
\email{klein@astron.berkeley.edu}

\begin{abstract}
Ambipolar diffusion is important in redistributing magnetic flux and in damping
\alfven waves in molecular clouds. The importance of
ambipolar diffusion on a length scale $\ell$ is governed by the
ambipolar diffusion Reynolds number,
$\rad=\ell/\lad$, where $\lad$ is the characteristic length scale for ambipolar diffusion.
The logarithmic mean of the AD Reynolds number in a sample of 15 molecular clumps with measured magnetic fields \citep{cru99} is 17, comparable to the 
theoretically expected value.  We identify several regimes of ambipolar diffusion
in a turbulent medium, 
depending on the ratio of the flow time to collision times between ions and neutrals; 
the clumps observed by \citet{cru99} are all in the 
standard regime of ambipolar diffusion, in which the neutrals and
ions are coupled over a flow time. 
We have carried out two-fluid simulations of 
ambipolar diffusion in isothermal, turbulent boxes for a range of values of $\rad$.
The mean Mach numbers were fixed at $\calm=3$ and $\ma=0.67$; self-gravity 
was not included. We study the properties of overdensities--i.e., clumps--in the simulation
and show that 
the slope of the higher-mass portion of the clump mass spectrum
increases as $\rad$ decreases, which is qualitatively
consistent with \citet{pad07}'s finding that the mass spectrum in hydrodynamic
turbulence is significantly steeper than in ideal MHD turbulence. 
For a value of $\rad$ similar to the observed value, we find a slope that is consistent
with that of the high-mass end of the Initial Mass Function for stars.
However, the value we find for the spectral index in our ideal MHD simulation differs
from theirs, presumably because our simulations have different initial conditions.
This suggests that the mass spectrum of the clumps in 
the \citet{pad07} turbulent fragmentation model
for the IMF depends on the environment,
which would conflict with evidence for a universal IMF.
In addition, 
we give a general discussion of how the results of simulations of magnetized, turbulent,
isothermal boxes can be scaled to physical systems.
Each physical process that is introduced into the simulation,
such as ambipolar diffusion, introduces a dimensionless parameter, such as $\rad$,
which must be fixed for the simulation, thereby reducing the number of scaling parameters
by one. We show that the importance of self-gravity is fixed in
any simulation of ambipolar diffusion;
it is not possible to carry out a simulation in which self-gravity and ambipolar diffusion 
are varied independently unless the ionization is a free parameter.
We show that our simulations apply to small regions in molecular clouds, generally
with $\ell_0\la 0.4$~pc and $M\la 25\;M_\odot$.
A general discussion of the scaling relations 
for magnetized, isothermal, turbulent boxes, including self-gravitating systems, is given in the Appendix.
\end{abstract}

\section{Introduction}

Giant molecular clouds, threaded by magnetic fields, are the birth places for new stars.  Since the earliest studies of star formation, it has been recognized that the magnetic flux in stars is many orders of magnitude less than that in the interstellar material from which the stars originated. \citet{mes56} suggested that ambipolar diffusion (AD) could resolve this problem by allowing magnetic flux to be redistributed during collapse due to the differential motion between the ionized and neutral gas. With effective shielding of high energy cosmic rays and radiation, the ionization fraction of gas inside high-density cloud cores can be $\leq 10^{-7}$ \citep[e.g.][]{cas98,ber99}, which renders AD efficient.  Star formation theory based on the AD-regulated, quasi-static collapse of molecular clouds \citep[e.g.][]{spi68,nak72,mou76,mou77,mou79,nak78,shu83,liz89,fie92,fie93} naturally accounts for the enormous loss of magnetic flux during star formation.  

	However, both observations \citep{zuce74,zucp74} and theory \citep{aro75} have long indicated that supersonic turbulent motions are important in molecular clouds, and this turbulence has a major effect on star formation \citep{mac04,bal07,mck07}.  The kinetic energy of the supersonic motions is observed to be comparable to the magnetic energy of the clouds, so that molecular clouds are in approximate equipartition \citep[e.g.][]{cru99,hei05,tro08}.  
It should be borne in mind that 
the amplitude of turbulent fluctuations decreases with decreasing scale;
for example,
\citet{goo98} and \citet{bar98} find that the NH$_{3}$ lines within $\sim0.1$ pc of the centers of the cores that they examined do not obey the line width-size relations as seen on the large scale \citep[e.g.][]{lar81,sol87,hey04}.  
\citet{mou87} and \citet{mye98} have
argued that AD would damp turbulent motions on small scales, and
\citet{goo98} suggest that this damping could be 
enhanced by the 
low ionization fraction in the dense inner regions of the cores. 
Better data will enable determination of the role of ambipolar diffusion on
the small scales relevant to the formation of individual stars. Existing
data clearly show that turbulence is important
on larger scales, but observational tests of the theoretical prediction that
turbulence can accelerate the rate of AD \citep{fat02,zwe02} will be challenging.

Numerical simulation is an important tool in understanding supersonic turbulence in magnetized MCs, but it is very challenging to carry out three-dimensional (3D) simulations 
that include ambipolar diffusion.  The small ionization fraction in molecular clouds means that the ion inertia can be neglected. This permits a single-fluid treatment of ambipolar diffusion, which gives the induction equation the form of a diffusion equation \citep[e.g.][]{mac95,duf08}. However, in this case the 
the stability condition for explicit codes requires the time step to scale as 
the square of the grid-size ($\Delta x^2$
---\citealp{mac95}), which is prohibitive at high resolution \citep[e.g.][]{nar08}.   
Li, McKee, \& Klein (2006; hereafter LMK) developed the Heavy-Ion Approximation, which takes advantage of the negligible ion inertia in regions of very low ionization and can accelerate simulations of ambipolar diffusion by large factors.  
In the Heavy-Ion Approximation, the mass-weighted ionization is increased
by a factor $\calr\sim 10^4$ and the ion-neutral coupling coefficient is decreased
by the same factor, so that the momentum transfer between ions and neutrals is
unaffected.
Using a semi-implicit two-fluid scheme proposed by \citet{mac97} \citep[see also][]{tot95}, LMK tested the Heavy-Ion Approximation with several classical problems involving ambipolar diffusion and found speed-ups of order a factor 100. 

In the first astrophysical application of the Heavy-Ion Approximation, 
Li et al (2008, hereafter LMKF) studied the statistical properties of supersonically turbulent systems with ambipolar diffusion. The properties of the turbulence were found to vary smoothly from the hydrodynamic case to the ideal MHD case as the importance of ambipolar diffusion decreased.  They
found that the power spectra for the neutral gas properties of a strongly magnetized medium with strong ambipolar diffusion are similar to those for a weakly magnetized medium; in particular,
the power spectrum for the neutral velocity is close to that for Burger's turbulence.

	In this paper, we extend this work on turbulent systems with ambipolar diffusion but 
without self-gravity. This paper has three main goals: First, we 
give a general discussion of the ambipolar diffusion Reynolds number, $\rad$, that
characterizes ambipolar diffusion \citep{mye95,zwe02} 
(\S 2). 
We then determine
the numerical values of $\rad$ for the molecular regions studied
by \citet{cru99} (\S 2) and show that they are consistent with
the theoretically expected ones. 
Second, we use numerical simulations
to determine the properties of the clumps that appear
in a turbulent medium with ambipolar diffusion (\S\S 3 \& 4). 
In particular, we show how the mass function and 
the mass-to-flux ratio of the clumps depend on $\rad$. 
Third, we analyze the scaling properties of simulations with
ambipolar diffusion and determine the range of physical parameters that characterize the 
simulations (\S 5). This discussion is continued in the Appendix, which gives a
general discussion of how the results of simulations of turbulent boxes
can be applied to physical systems, including those that are
self-gravitating.
Further results from these simulations, particularly those relevant to measuring
the strength of the magnetic field
and determining the effects of heating due to ambipolar diffusion, 
will be discussed in a future paper (Paper III).

\section{The Ambipolar Diffusion Reynolds Number}
\label{sec:AD}

	The effects of ambipolar diffusion on a length scale $\ell$ 
in a medium with a flow velocity $v$
can be characterized by the ambipolar diffusion (AD) Reynolds number, $R_{\rm AD}(\ell)$.
This quantity appears to have been first introduced by \citet{mye95};
they referred to it as the magnetic Reynolds number, although that
term is normally used to describe the effects of Ohmic resistivity.
The AD Reynolds number 
is motivated as follows \citep[][LMK]{zwe97,zwe02}:
Ions in a partially ionized plasma are subject to two forces: 
the Lorentz force, $\sim \brms^2/4\pi\ell_B$,
where $\brms$ is the rms magnetic field strength and
$\ell_B\equiv|\brms/\nabla \brms|$; and the drag force, $\gad\rho_i\rho_n\vad$,
where $\gad$ is the ion-neutral coupling coefficient, $\rho_i$ and $\rho_n$
are the ion and neutral densities, respectively, and $\vad$ is the drift velocity
between the neutrals and the ions. When the ionization is low enough that the
ion inertia can be neglected, these forces balance and the drift velocity is
\beq
\vad(\ell_B)\simeq \frac{\brms^2}{4\pi\gad\rho_i\rho_n\ell_B}.
\eeq
We define the ambipolar-diffusion time over a length scale $\ell$ as
\beq
\tad(\ell)\equiv\frac{\ell}{\vad(\ell)}=\frac{4\pi\gad\rho_i\rho_n\ell^2}{\brms^2}.
\eeq
Similarly, we can introduce the ambipolar-diffusion length scale $\lad$, which is
the length for which the ambipolar drift velocity is the same as the flow velocity---i.e.,
in the frame of the ions, the length scale over which the field varies in a steady flow:
\beq
\lad=\frac{\brms^2}{4\pi\gad\rho_i\rho_n v}.
\eeq
In terms of the neutral-ion collision time, $t_{ni}=1/\gad\rho_i$, the AD time scale and
length scale are
\beqa
\tad&=&\frac{\ell^2}{\va^2t_{ni}},\\
\lad&=&\frac{\va^2t_{ni}}{v},
\eeqa
where $\va=\brms/(4\pi\rho)^{1/2}$ is the \alfven velocity and
where we have assumed that the ion mass density is negligible, so that $\rho_n\simeq \rho$.
The effect of ambipolar diffusion on a flow over a length scale $\ell$ with
a characteristic
velocity $v$ is determined by
the AD Reynolds number,
\beq
R_{\rm AD}(\ell)\equiv\frac{\ell v}{v_A^2 t_{ni}}=
\frac{\tad}{t_f}=
\frac{\ell}{\lad}=
\frac{4\pi\gad\bar\rho_i\bar\rho_n \ell v}{\brms^2},
\label{eq:radell}
\eeq
where $t_f\equiv \ell/v$ is the flow time across a length $\ell$.
Observe that ambipolar 
diffusion increases in importance as $\rad(\ell)$ decreases; thus, it becomes more
important at low densities, low ionizations,
low velocities, small distances and high field strengths.
As \citet{mye95} showed, the ratio of the size of the region, $\ell$,
to the minimum wavelength of a propagating \alfven wave in which the
inertia is provided by the neutrals ,
$\lambda_{\rm min}=\pi \va t_{ni}$ \citep{kul69}, is directly proportional to $\rad(\ell)$:
\beq
\frac{\ell}{\lambda_{\rm min}}=\frac{\rad(\ell)}{\pi \ma},
\label{eq:lammin}
\eeq
where $\ma\equiv v/\va$ is the \alfven Mach number.

We have defined the AD Reynolds number in terms of the mean densities and the rms field
strength. In a supersonically turbulent medium, the densities are subject to large fluctuations,
and if the \alfven Mach number
is large also, the magnetic field has large fluctuations as well.
If one defines $\rad$ in terms of the local densities and field strength, then
one can devise several different ways of averaging so as to obtain an
effective value of $\rad$ for a turbulent medium; in particular, the length
scale $\ell$ can be taken to be the size of the region or it can be determined
self-consistently as the size of the average eddy or density fluctuation.
The resulting values for $\rad$ in a turbulent region of size $\ell_0$ range from slightly larger
than $\radlo$ to several times less (see Paper III for further discussion).
One should thus bear in mind that $\radlo$ is a characteristic value for
the ratio of the ambipolar diffusion time to the flow time, and the actual
value in a turbulent medium might differ from this by a factor of a few.

Mouschovias (private communication) has emphasized that the AD Reynolds number
is useful for turbulent media in which the velocity dispersion is 
determined by the turbulence (the case we
are considering here), but not in systems in which the flow velocity is
determined by the AD process itself. For example, in 
quasi-static, AD-regulated star formation, the AD length
scale, $\lad$ is
proportional to the radius of the self-gravitating cloud, and $\rad$ is of order
unity. Similarly, $\rad$ is not a useful parameter to characterize C-shocks
\citep{dra80}, since the structure of such shocks adjusts itself so that $\rad\sim 1$
\citep{li06}.

\subsection{Numerical Evaluation of $\radl$}

	Evaluation of the AD Reynolds number requires evaluation of both
the ion-neutral coupling coefficient, $\gad$, and of the mean ionization
mass fraction,
$\bar\chi_i\equiv \bar\rho_i/\bar\rho$. If
this mass fraction is small ($\bar\chi_i\ll 1$), then $\bar\rho_n\simeq \bar\rho$ and
\beq
\radl=\frac 12\left(\frac{\gad\bar\rho \ell}{c_s}\right)\bar\chi_i\calm\beta
=\left(\frac{\gad\bar\rho \ell}{c_s}\right)\frac{\bar\chi_i\ma^2}{\calm},
\label{eq:radltwo}
\eeq
where 
$c_s$ is the isothermal sound speed, $\calm\equiv v/c_s$ is the Mach number and
$\beta\equiv 8\pi\bar\rho c_s^2/\brms^2$ is
the plasma $\beta$ parameter.
We normalize our results to the case in which the ionization is dominated by
HCO$^+$. The ion-neutral coupling coefficient is then
\beq
\gad=\frac{1.9\times 10^{-9}\;\mbox{cm$^3$ s$^{-1}$}}{m_n+m_i}=3.7\times 10^{13}~~
	\mbox{cm$^3$ s$^{-1}$ g$^{-1}$}
\eeq
\citep{dra83}, provided the relative velocity of the ions and neutrals is less than
about 19 km s\e. 
Note that this value of the coupling coefficient differs from that adopted in LMKF due to
our assumption that the ionization is dominated by HCO$^+$.
More generally, we shall write
\beq
\gad=3.7\times 10^{13}\gad^*~~
	\mbox{cm$^3$ s$^{-1}$ g$^{-1}$},
\label{eq:gads}
\eeq
where $\gad^*$ is a number that allows for ions other than HCO$^+$.

	Next, we consider the ionization. 
The processes that determine the ionization in molecular clouds are complex, 
and in general the ionization is time-dependent. We adopt a characteristic value
of the ionization based on the assumption that the ionization is in a steady state
and is dominated by HCO$^+$. In equilibrium, the mean ionization fraction by number
is
\beq
\bar \xeeq=\left(\frac{\zeta_{\rm CR}}{\alpha\bar\nh}\right)^{1/2},
\label{eq:xeeq}
\eeq
where $\nbh=\bar\rho/\muh$ is the mean density of H nuclei,
$\muh=2.34\times 10^{-24}$~g is the mass per H nucleus,
$\zeta_{\rm CR}\sim 3\times 10^{-17}$~s\e\ is the cosmic ray ionization rate
per H atom \citep[see the discussion by][]{dal06}
and $\alpha$ is the relevant recombination rate \citep{mck07}.
Equation (\ref{eq:xeeq}) is consistent with the results of \citet{pad04} at late
times and high densities for
$\alpha=2.5\times 10^{-6}$~cm$^3$~s\e, the value they adopted for
the dissociative recombination rate of HCO$^+$. 
If small PAHs dominate the ionization, then the dissociative recombination
rate is about 10 times smaller \citep{wak08} and the ionization several
times larger.
\citet{tas05,tas07}, who included the effects
of charged grains, adopted a dissociative recombination rate
$\alpha=1.0\times 10^{-6}$~cm$^3$~s\e; their results for the ionization are
approximately consistent with equation (\ref{eq:xeeq}) for densities $\nh\la 10^8$~cm\eee.

Inference of the ionization from observations generally requires knowledge of
the cosmic ray ionization rate and the density as inputs to the
chemical models used to interpret the observations
\citep{wil98,pad04}. The ionization can be characterized
by the parameter $C_i$ defined by
\beq
x_e\equiv C_i\left(\frac{\zcr}{\nh}\right)^{1/2}.
\eeq
In the equilibrium model above, $C_i=\alpha^{-1/2}$, which is 630~cm$^{3/2}$~s$^{1/2}$
for the fiducial case. \citet{wil98} found a median ionization 
$x_e=4.5\times 10^{-8}$ (note that they normalized their results to \htwo, whereas
we are normalizing with respect to H). Their adopted ionization
rate ($\zcr=2.5\times 10^{-17}$~s\e) and density ($\nh=5\times 10^4$~cm\eee)
correspond to $C_i\simeq 2000$~cm$^{3/2}$~s$^{1/2}$. More recently, 
\citet{pad04} have interpreted these data with time dependent models
and infer lower values of the ionization and therefore $C_i$. They find
$x_e\simeq 7.5\times 10^{-9} - 3.5\times 10^{-8}$ and attribute the
higher values to the effect of FUV photoionization; for their assumed
ionization rate ($\zcr=6\times 10^{-18}$~s\e) and density ($2\times 10^4$~cm\eee),
the implied value of the ionization parameter is $C_i\sim 600$~cm$^{3/2}$~s$^{1/2}$,
fortuitously close to our fiducial value. The difference between the
values of $C_i$ inferred by \citet{wil98} and \citet{pad04} is a reflection
of the uncertainties that remain in determining the ionization in molecular clouds.

	To evaluate the AD Reynolds number, we require 
the ion {\it mass} fraction $\bar\chi_i$, which is related to the ion number
fraction $\bar x_i\equiv \bar n_i/\bar\nh$ by $\bar\chi_i=\bar x_i m_i/\muh
\rightarrow 20.7 \bar x_i$, where
the numerical evaluation is for HCO$^+$. We then have
\beq
\bar\chi_i=\frac{m_i C_i}{\muh}\left(\frac{\zeta_{\rm CR}}{\bar\nh}\right)^{1/2}
=2.25\times 10^{-6}\left(\frac{\chiis}{\nbht^{1/2}}\right),
\label{eq:chii}
\eeq
where the numerical factor
\beq
\chiis=\frac{m_i}{29\;\mbox{amu}}\left(\frac{C_i}{630~\mbox{cm}^{3/2}~\mbox{s}^{1/2}}
\right)\left(\frac{\zcr}{3\times 10^{-17}~\mbox{s}^{-1}}\right)^{1/2}
\label{eq:chiis}
\eeq
allows for deviations from the fiducial case
and where $\nbht\equiv\nbh/(10^3$~cm\eee). 
Under the assumption that the ion mass is indeed about
29~amu (i.e., the mass of HCO$^+$), the results of \citet{wil98} correspond to $\chiis\simeq 3$,
whereas those of \citep{pad04} correspond to $\chiis\simeq 1$.
The ionization 
can also be expressed in the form $\rho_i=\calc\rho^{1/2}$ (Shu 1983),
with
$\calc=\chi_i\rho^{1/2}=
1.09
\times 10^{-16}\chiis$~g$^{1/2}$~cm$^{-3/2}$.
Shu (1983) adopted a value $\calc=3\times 10^{-16}$~g$^{1/2}$, corresponding
to $\chiis\sim 3$, in agreement with the estimate of \citet{wil98}.
Numerically, the AD Reynolds number is then
\beq
R_{\rm AD}(\ell)=
16.0
\gad^*\chiis\calm\beta\left(\frac{\nbht^{1/2}\ell_{\rm pc}}{T_1^{1/2}} \right),
\label{eq:radladtext}
\eeq
where $T_1\equiv T/(10$~K) and $\ell_\pc\equiv\ell/$(1~pc).

\subsection{Regimes of Ambipolar Diffusion}
\label{sec:regimes}

	We can distinguish several regimes in ambipolar diffusion in a 
turbulent medium. 
For $\chi_i\equiv\rho_i/\rho\ll 1$,
we have $\rho_n\simeq\rho$ so that
the neutral-ion collision time, $t_{ni}$ and the corresponding
ion-neutral collision time, $t_{in}$ are related by
\beq
t_{in}\equiv\frac{1}{\gad\rho_n}=\frac{\chi_i}{\gad\rho_i}=\chi_i t_{ni}.
\eeq
Similarly, the ion-\alfven Mach number,
$\mai\equiv v(4\pi\rho_i)^{1/2}/\brms$, is related to the \alfven Mach number, $\ma$, 
by $\mai^2=\chi_i\ma^2$. It follows that
\beq
\rad=\ma^2\left(\frac{t_f}{t_{ni}}\right)=\mai^2\left(\frac{t_f}{t_{in}}\right).
\eeq

	We now identify five different regimes for ambipolar diffusion. For simplicity
we ignore possible differences between the velocity dispersions of the neutrals and ions
(to be discussed in Paper III), which
could change the coefficient in front of $\mai^2$ by up to a factor 2 in the expressions below.

\begin{itemize}
\item[I.] Ideal MHD ($t_f/t_{ni}\rightarrow\infty$, corresponding to
$\rad\rightarrow\infty$ for a given value of $\ma$): 
The ions and neutrals are
perfectly coupled.

\item[II.] Standard AD ($t_f>t_{ni}\gg t_{in}$, corresponding to $\rad>\ma^2$): 
The neutrals and ions are coupled together over a flow time
so that the AD is weak.
For $\ma={\cal{O}}(1)$, linear
\alfven waves can propagate, since the propagation condition 
for \alfven waves of wavelength $\ell$ derived by
\cite{kul69} is equivalent to $\rad(\ell)>\pi\ma$ (eq. \ref{eq:lammin}). The wave damping is weak
(i.e., $\Gamma t_f<1$,
where $\Gamma=\frac 12 k^2\va^2t_{ni}$ is 
the damping rate for low-frequency waves) for $\rad(\ell)>2\pi^2$.

\item[III.] Strong AD ($t_{ni}>t_f>t_{in}$, corresponding to $\ma^2>\rad>\mai^2$):
The neutrals are no longer coupled to the ions in a flow time, but the ions
remain coupled to the neutrals.
For $\ma={\cal{O}}(1)$, \alfven waves cannot propagate since $\lambda_{\rm min}/\ell
>\pi/\ma\ga 1$ (eq. \ref{eq:lammin}).

\item[IV.] Weakly coupled ($t_{in}>t_f$, corresponding to $\mai^2>\rad$): 
The ions and neutrals are only weakly coupled and act almost independently.
The damping rate for the high-frequency \alfven waves that propagate in the ions
is $\Gamma=1/(2t_{in})$, so these waves are weakly damped in this regime:
$\Gamma t_f=\rad(\ell)/(2\mai^2)<1$ for waves of wavelength $\ell$. 
The Heavy-Ion Approximation is based on
the assumption that the ion inertia is negligible and therefore does not apply
to this regime (see below).

\item[V.] Hydrodynamics 
($t_f/t_{in}\rightarrow 0$ or $\chi_i\rightarrow 0$, corresponding to $\rad\rightarrow 0$
for a given value of $\ma$): The neutrals are not affected by
the trace ions and act purely hydrodynamically.
One can of course recover the hydrodynamic limit by letting $B\rightarrow 0$
so that $\ma\rightarrow\infty$; in that case, $\rad$ is unconstrained.
It should be noted that the boundary between the hydrodynamic 
regime and the weakly coupled 
regime is a matter of choice; if one demands that the ions have at most a
1\% effect on the neutrals, for example, then $\rad$ would have to be smaller
than if one demands that the effects be limited to 10\%.

\end{itemize}
\noindent
It should be borne in mind that in all regimes except the last (where it is irrelevant),
we have assumed that
the ions are well-coupled to the magnetic field---i.e., the ion gyrofrequency is
much larger than the ion-neutral collision frequency, $\Omega_it_{in}\gg 1$.
Although we have defined the AD regimes for arbitrary values of the
\alfven Mach number (provided $B$ is large enough that $\Omega_it_{in}\gg 1$),
this characterization of ambipolar diffusion is most useful when $\ma\sim
{\cal{O}}(1)$, as it generally is in molecular gas in the interstellar medium.

\subsection{Observed Values of $\rad$}
\label{sec:obs}

\citet{cru99} has summarized sensitive Zeeman measurements of magnetic field strengths together with other physical parameters, including the plasma $\beta$ and the Mach numbers, 
for 27 
molecular clouds. Of these, 12 have only an upper limit on the line-of-sight magnetic field.  Table 1  lists the values of the parameters from Tables 1 and 2 in \citet{cru99} that we use to compute the corresponding $\rad$ using equation (\ref{eq:radladtext}). 
We take the length scale $\ell_0$ to be the cloud diameter. 
We use Crutcher's correction for projection effects on the magnetic field: Zeeman
observations determine the line-of-sight component of the field, $B_{\rm los}$, and on average
the value of $B^2$ that enters the plasma-$\beta$ parameter is $3 B_{\rm los}^2$.
We assume that the parameters describing the ion-neutral coupling and the ionization
($\gad^*$ and $\chiis$) are unity.
From Table 1, we see that clouds with measured field strengths have $\rad$ ranging from a few to $\sim 70$.
Because the range of $\rad$ is so large, we quote the logarithmic average,
defined as 
\beq
\avg{\rad}_{\rm log}\equiv 10^{\avg{\log\rad}};
\eeq
 the logarithmic mean and dispersion of the AD Reynolds number in these clouds
is $\avg{\rad}_{\log} = 17\pm0.4$~dex.  
Clouds that have only upper limits on the magnetic field have an average
lower limit on the AD Reynolds number of $\avg{\rad}_{\log}=22$; if we discard
L889 as an outlier because of its unusually high Mach number ($\calm=7.3$), the
logarithmic mean is $18$, which is comparable to that of the clouds with measured fields.
We also include the \alfven Mach number, $\calm_{\rm A}$, in Table 1.  All the clouds have $\rad>\ma^2$, implying that these clouds are in the standard AD regime
(\S \ref{sec:regimes}).  
The ratio of the size of the cloud to the minimum \alfven wavelength is in the
range $2-15$ for clouds with measured field strengths.
We conclude that the effects of ambipolar diffusion must be considered in studies of 
molecular clouds, at least in those regions shielded from the interstellar
radiation field so that $\chiis={\cal{O}}(1)$,
in agreement with studies extending back for many years (e.g., \citealp{mou87}).

\subsection{Predicted $\rad$ and Implied Self-Gravity}
\label{sec:adsg}

	 As we now show, it is possible to predict the AD Reynolds number
for self-gravitating clouds that have an ionization of the form given in equation
(\ref{eq:chii}).
As a corollary, we show that the strength of self-gravity is
not a free parameter in simulations of
ambipolar diffusion in a turbulent medium.

	The importance of self-gravity in a cloud 
of radius $R_0$ or in a simulation box of size $\ell_0=2R_0$ is determined by
the virial parameter \citep{ber92}, 
\beq
\avir\equiv\frac{5\sigma^2 R_0}{G M_0}=\frac{5\sigma^2 \ell_0}{2G M_0},
\label{eq:avir1}
\eeq
where
\beq
\sigma=\frac{1}{\surd 3}\;\calm c_s
\eeq 
is the 1D velocity dispersion in the cloud.
The virial parameter is thus proportional to the ratio of
kinetic to gravitational energy.
We wish to treat both real clouds, which we approximate as effectively
spherical, and  turbulent boxes.
Of course, real clouds are not spherical (\citealp{ber92} give the generalization
to elliptical clouds), but keeping track of these two cases provides a gauge of
the importance of geometric effects; furthermore, the spherical cloud model
has long been in use (e.g., \citealp{sol87}). 
Let the area  and volume of the cloud or box be 
\beqa
A&\equiv& c_A \ell_0^2,
\label{eq:ca}\\
V&\equiv& c_V\ell_0^3,
\label{eq:cv}
\eeqa 
where $c_A=(\pi/4, 1)$ and
$c_V=(\pi/6,1)$ for a spherical cloud and a box, respectively.
The virial parameter then becomes
\beq
\avir=\frac{5 \calm^2 c_s^2}{6c_V G\bar\rho \ell_0^2}.
\label{eq:avir2}
\eeq
Since $\bar\rho_i\equiv\bar\chi_i\bar\rho$, equation (\ref{eq:radltwo}) 
for the AD Reynolds number can be
rewritten as
\beq
\radlo=\left(\frac{5}{6c_V G\avir}\right)^{1/2}\gad \bar\chi_i\bar\rho^{1/2}\ma^2,
\eeq
which  shows that the AD Reynolds number is determined by the ionization,
the \alfven Mach number and the virial parameter. Insofar
as the ionization is a function of the density, $\radlo$ will
also depend on density. However, in the case of greatest interest, in which
$\bar\chi_i\propto \chiis\bar\rho^{-1/2}$, where $\chiis$ is a number
that is unity in the fiducial case (eq. \ref{eq:chiis}),
the AD Reynolds number is fixed at
\beq
\radlo=19.7\gad^*\chiis\left(\frac{\pi}{6c_V \avir}\right)^{1/2}\ma^2.
\label{eq:radlocloud}
\eeq

Molecular cloud cores and clumps with measured magnetic fields are typically 
self-gravitating, with $\avir\sim 1$, and have $\ma\sim 1$ (Crutcher 1999).
As a result, for the fiducial values of $\gad$ and $\chi_i$, such regions have $\radl\sim 20$. 
This predicted value is in good agreement with the observed values discussed in
\S \ref{sec:obs}, which have a logarithmic mean of 17 and a dispersion of 0.4 dex.
GMCs as a whole have larger values of $\radl$ since most of their mass is photoionized by UV radiation \citep{mck89}, so that they have a higher ionization  than the cores and clumps within them (e.g., in an envelope of a GMC in which the ionization is dominated
by C$^+$, the ionization parameter is $\chiis\sim 10^2$).

The importance of self-gravity in a magnetized medium can also be expressed
in terms of the ratio of the mass to the magnetic critical
mass, $M_\Phi$, which is the minimum mass that can undergo gravitational collapse.
In terms of the magnetic flux,
$\Phi\equiv B c_A \ell_0^2$,
the magnetic critical mass is
\beq
M_\Phi=c_\Phi\frac{\Phi}{G^{1/2}}\ ,
\label{eq:mphi}
\eeq
where $c_\Phi = 1/2\pi$ for a cold sheet \citep{nak78} and $\approx 0.12$ for a cloud with a flux-to-mass distribution corresponding to a uniform field threading a uniform spherical cloud \citep{mousp76,tom88}.
For $c_\Phi=1/2\pi$, the ratio of the mass to the magnetic critical mass is
\beqa
\mu_{\Phi,\,0}\equiv\frac{ M_0}{M_\Phi}
	&=&\left(\frac{5\pi c_V}{6c_A^2 \avir}\right)^{1/2}\ma
		\label{eq:muphi}\\
&\rightarrow& \left(\frac{20}{9}\right)^{1/2}\frac{\ma}{\avir^{1/2}}~~~~~~\mbox{(spherical cloud)},
\label{eq:muphis}
\eeqa
which provides a simple relation between the \alfven Mach number, $\ma$ and
the two parameters describing the importance of self gravity in a magnetized,
turbulent cloud, $\avir$ and $\mu_{\Phi,\,0}$.
The ratio
$\mu_{\Phi,0}$ is sometimes written as the ratio of the observed
mass-to-flux ratio to the critical one, $(M/\Phi)_{\rm obs}/(M/\Phi)_{\rm crit}$ 
(e.g., \citealp{tro08}).
Using equation (\ref{eq:radlocloud}), we find that the AD Reynolds number is
given in terms of $\mu_{\Phi,\,0}$ by
\beq
\radlo=13.2\left(\frac{2c_A}{3c_V}\right)\gad^*\chiis\ma\mu_{\Phi,\,0};
\eeq
the factor in parentheses is unity for a spherical cloud.
Gravitationally bound clouds that are both
magnetized and turbulent have $\mu_{\Phi,\,0}$ somewhat
greater than unity since the gravity has to overcome both the turbulent
motions and the magnetic field \citep{mck89}. This expression thus gives
a similar result to that in equation (\ref{eq:radlocloud}) for $\ma\sim 1$
and $\mu_{\phi,\,0}\simeq 1-2$.

These relations for $\radlo$ can be inverted to give the values of
the virial parameter and the ratio of the mass to the critical mass
in terms of $\radlo$ and $\ma$. In other words, a simulation of
a turbulent box with ambipolar diffusion [which requires
specification of $\radlo$ and $\ma$] necessarily implies the
strength self-gravity would have were it to be included:
\beqa
\avir&=& 203\left[\frac{\gad^*\chiis}{\radlo}\right]^2\ma^4\; ,
\label{eq:avirtext}\\
\mu_{\Phi,\,0}&=& 0.114 \left[\frac{\radlo}{\gad^*\chiis}\right]\frac{1}{\ma}\; ,
\label{eq:muphisim}
\eeqa
where we have set $c_A=c_V=1$, as is appropriate for a simulation box.
For $\avir\gg 1$, the neglect of self-gravity is
self-consistent. Parameter choices that lead to values of $\avir\ll 1$ 
and $\mu_{\Phi,\,0}\ga 1$
are not self-consistent, since self-gravity would lead to
turbulent motions that render $\avir\ga 1$
(e.g., \citealp{kle09}).

We emphasize that ``implied self gravity" does not mean that simulations of
ambipolar diffusion mimic the effects of self gravity. Rather, it means that
the strength of the self-gravity, were it to be included, is not a free parameter
provided the ionization parameter $\chiis$ is specified.\footnote{
Of course, the relation between $\avir$ and $\radlo$ also depends on $\ma$,
but this is fixed in simulations of turbulent boxes with $\ma\la 1$; by contrast,
whereas observed clouds have definite values of $\chiis$, it is not necessary
to specify this quantity in the simulation---see \S \ref{sec:codeunits}.}
By contrast, a simulation of a turbulent box with ideal MHD is scale free;
the density can be chosen so that self-gravity would be negligible if it
were included. This freedom does not exist in simulations of ambipolar diffusion.

\section{Simulations}

In this paper, we extend the LMKF study of supersonic turbulence with ambipolar diffusion, focusing on the physical properties of the clumps formed purely as the result of turbulent fragmentation with no gravity.  LMKF
performed a series of $256^3$ simulations 
in a periodic box using the code ZEUS-MPAD to investigate turbulence statistics in non-ideal MHD without self-gravity. 
Like LMKF, we drove 
the turbulence with a fixed driving pattern 
over the wavenumber range $1\leq k\leq 2$ 
(where $k\equiv \kphys\ell_0/2\pi=\ell_0/\lambda$ is the normalized wavenumber)
using the recipe described in \citet{mac99}.   
The driving maintained the 3D Mach number at $\calm=3$,
which is only mildly supersonic.
The corresponding line-of-sight Mach number---i.e.,
the 1D Mach number $\calm/\surd 3$---is less than 2.
The magnetic field was initially uniform, with a strength set by
a plasma-$\beta$ parameter of 0.1, 
corresponding to an \alfven Mach number
$\ma=0.67$; the turbulence is thus sub-\alfvenicstop.
During the simulations, the volume-averaged magnetic field changed by less
than 10\%, and as a result the volume-averaged value of $\beta$ remained within
10\% of its initial value.
As shown in Table 1, this value of $\beta$ is close to the median of the 15 clouds 
with measured magnetic fields.

The focus of our effort is to determine how the properties of the clumps vary with $\radlo$,
so for now we discuss our results in dimensionless form;
the physical conditions corresponding to these
simulations will be discussed in \S \ref{sec:scaling} below.
We note, however, that for systems satisfying the linewidth-size relation,
a Mach number of 3 corresponds to a box size $\ell_0\simeq 0.4$~pc.
Like LMKF, we considered values of $\radlo$ from 0.12, close to the hydrodynamic limit, to
1200, close to the ideal MHD limit. The run with $\radlo=12$ has
conditions similar to those in observed clouds; as we shall see below,
if we assume that the simulated region satisfies the linewidth-size relation,
its density would be $\nbh\simeq 10^4$~cm\eee. 
In \S \ref{sec:sonic}, we show that the inertial range of the simulated turbulence extends
over the range $\ell_0/3 - \ell_0/20$, so the AD length scale $\lad=\ell_0/\radlo$
is in the inertial range for this run. For all the other runs, the AD length scale
is outside the inertial range.
The run with $\radlo=1.2$ 
focuses on scales less than $\lad$ and
has a lower
value of the AD Reynolds number than any of the clouds observed by
\citet{cru99}, most of which are gravitationally bound. 
If the simulation satisfied the linewidth-size relation,
it would have a density of $\nbh\simeq  400$~cm\eee,
corresponding to an unbound cloud. 
On the other hand, the runs with $\radlo=120,\, 1200$
focus on scales greater than $\lad$ and
have higher AD Reynolds numbers than any of the clouds with measured
magnetic fields in that sample. The $\radlo=1200$ run could
be applied to the outer parts of GMCs, where the ionization is
dominated by C$^+$. The
run with $\radlo=0.12$ represents the transition to the hydrodynamic
limit, and is 
primarily of theoretical
rather than practical interest.

All models were run for $3t_f$, where $t_f\equiv \ell_0/\calm c_s$ is the flow time. 
In order to improve the statistics and the resolution, we re-ran the five models m3c2r-1
[$\radlo=0.12$] to m3c2r3 [$\radlo=1200$; 
see Table 2] in LMKF with the same initial conditions but using a $512^3$ grid.  
All the results reported in this paper are the result of simulations on such a grid.
The total computing time for all the models was $\sim$ 600,000 CPU hours on the NCSA machine Abe using 512 processors. 

	We made two principal approximations in our simulations. First, as discussed in the Introduction, we used the Heavy Ion Approximation (LMK), 
adopting an ionization $\tilde\chio=\calr\chio$ and a corresponding
ion-neutral coupling coefficient $\tilde\gamma_{\rm AD}
\propto\gad/\calr$, with $\calr\sim 10^4$; here the tilde denotes
quantities measured in code units (see \S \ref{sec:codeunits}). The key to the Heavy Ion Approximation is that even though each of these parameters differs
from the actual value by a factor of $10^4$, the ion-neutral coupling
is governed by the product of the parameters and has the correct physical value.  
According to the discussion in \S 2.2, the five AD models and the ideal MHD model span three regimes of AD as listed in Table 2, based on the initial $\radlo$
in equilibrium.

	Our second principal approximation was in our treatment of the ionization.
Simulations can be carried out with various assumptions about the ionization,
including ion conservation, ionization equilibrium and time-dependent ionization.  Following LMKF, we assumed that the number of ions is conserved, so that the value of
$\bar\chi_i$ for the entire box is constant. The density is initially uniform,
so that the initial ionization mass
fraction, $\bar\chio$, is the same everywhere; we took
it to be $10^{-6}$.
LMKF demonstrated that the results were the same as
in the case of ionization equilibrium (basically because the time for a neutral to
exchange momentum with an ion is small compared to the ionization time scale).
More generally, the ionization is
time-dependent. The ratio of the flow time, $t_f\equiv \ell_0/\calm c_s$, 
to the characteristic
ionization time, $t_{\rm ion,\,eq}=\xeeq/\zcr$ (see eq. \ref{eq:xeeq}), is large:
\beqa
\frac{t_f}{t_{\rm ion,\,eq}}&=&(\alpha\zcr\nbh)^{1/2}
\left(\frac{\ell_0}{\calm c_s}\right),\\
&=&1.40\times 10^3(\alpha^*\zcr^*)^{1/2}\left[\frac{\radlo}{\gad^*\chiis}\right]
\frac{1}{\ma^2},
\eeqa
where $\alpha^*\equiv\alpha/(2.5\times 10^{-6}~$cm$^3$~s\e),
$\zcr^*\equiv\zcr/(3\times 10^{-17}~$s\e), and we have used equation (\ref{eq:radladtext}).
It follows that the molecular gas
is typically very close to ionization equilibrium (although it is not necessarily
close to chemical equilibrium). In simulations, the relevant comparison is between
the flow time across a cell, $t_f/\caln_g$, where $\caln_g$ is
the number of grid cells in the length of the box, and the ionization time.
For our runs, which typically have $\caln_g=512$, we have 
$t_f/(\caln_g t_{\rm ion,\,eq})\simeq 6\radlo$
for fiducial values of the parameters. Ionization equilibrium is thus a 
good approximation for all the cases we consider except $\radlo=0.12$.

	To test our use of the approximation of ion conservation \citep[see also the Appendix in][]{li08}, we ran several $256^3$ models with time-dependent ionization for 
different values of $\rad$.
We find that the properties of the clumps in
these runs are within a few percent of those in the corresponding $256^3$ runs with ion conservation, with the exception of the ion density. In fact, the mean ion
density in the entire box in the time-dependent case is less than that in the
ion conservation case by up to a factor $\sim 2$. As a result, the value
of the AD Reynolds number is reduced by a corresponding factor,
as shown in Table 2. For large $\radlo$, the gas is close to ionization
equilibrium, so that $\bar\rho_i\propto\rho^{1/2}$.
With this relation for the ion density, the mean ion density, and hence $\radlo$,
are reduced by only a small amount compared to the case of ion conservation
for the low Mach number we are considering if the density PDF is a lognormal with a
width similar to that found by \citet{pad02}. 
For small $\radlo$
the deviations from ionization equilibrium are larger, and correspondingly the
difference between the time-dependent and ion conservation results are larger as well.
In this paper, however, we are exploring the effects of changing $\radlo$ by
orders of magnitude, so changes of $\la 2$ do not affect our conclusions.

\section{Physical Properties of Clumps}

The formation of high-density clumps is a natural outcome in simulations of highly supersonic turbulence, whether a magnetic field is included or not. Furthermore, high-resolution turbulence simulations \citep[e.g.][]{li04,pad07} produce a mass spectrum of clumps that qualitatively resembles the stellar initial mass function (IMF), with a peak 
at low mass and a power-law tail at high masses. 
Recent observations of molecular cores \citep[e.g.][]{tac02,oni02,alv07} suggest a similarity between the stellar IMF and the core mass function. (We follow the terminology of \citealp{wil00} and use the term ``core" to refer to the subset of clumps that are gravitationally bound and will form a star or small multiple stellar system.)  \citet{pad02} and \citet{pad07} have proposed a turbulent fragmentation theory for the IMF that relates the index of the velocity power spectrum  to the slope of the higher-mass
end of the clump mass spectrum.  LMKF showed that ambipolar diffusion changes the velocity power index, and we confirm that conclusion in Paper III.
If the turbulent fragmentation theory is correct, we would expect a change in the slope of the higher-mass end of the clump mass distribution between the ideal MHD and the AD turbulence simulations as well.
(It should be noted that the \citealt{hen08} theory leads to a much smaller predicted difference
in the slope of the IMF in these two cases.)

We use a CLUMPFIND algorithm, based on the algorithm developed by \citet{wil94}, to determine the clumps in our simulations.  We define ``clumps" as connected regions with a density larger than the mean density of the turbulent box and will use the term ``\clmf" for ``clump mass function," reserving ``CMF" for ``core mass function."
This distinction is appropriate for our simulations since they do not include self gravity.
The density contours are separated by $\delta \rho = 0.04 \rho$,
which \citet{pad07} found to work well in distinguishing distinct clumps.
In order to infer the effects of AD on the \clmf, we require the clumps to be resolved.  As mentioned in LMK, ZEUS-MPAD needs at least 3 to 6 zones to accurately distinguish the effects of AD from those of numerical diffusion.  Therefore, we require clumps to have at least 6 zones in the mean radius,
unless otherwise specified; this requirement 
is validated in the resolution study of \clmf\  in \S5.2.1.
In implementing this resolution requirement, we define the effective radius as
$r_c\equiv (3V_c/4\pi)^{1/3}$, where $V_c$ is the volume of the clump is determined
by summing the volumes of each cell in the clump that has a density above
threshold; thus, for a porous clump, $r_c$ is less than the projected radius of the clump
(see \S \ref{sec:effect}). This approach to setting the resolution requirement eliminates
small, very porous clumps, which have a lot of structure that is not well resolved.
By varying $\delta \rho$, we found that the number of clumps with mean radius larger than 6 cells does not change when the separation of the density contours is smaller than 4\%,
thereby justifying our choice of $\delta\rho$.
Before constructing the \clmf, we verify that the clumps defined in our simulations satisfy the heavy-ion approximation.

\subsection{The Heavy-Ion Approximation for Clumps}

\begin{figure}
\epsscale{.80}
\plotone{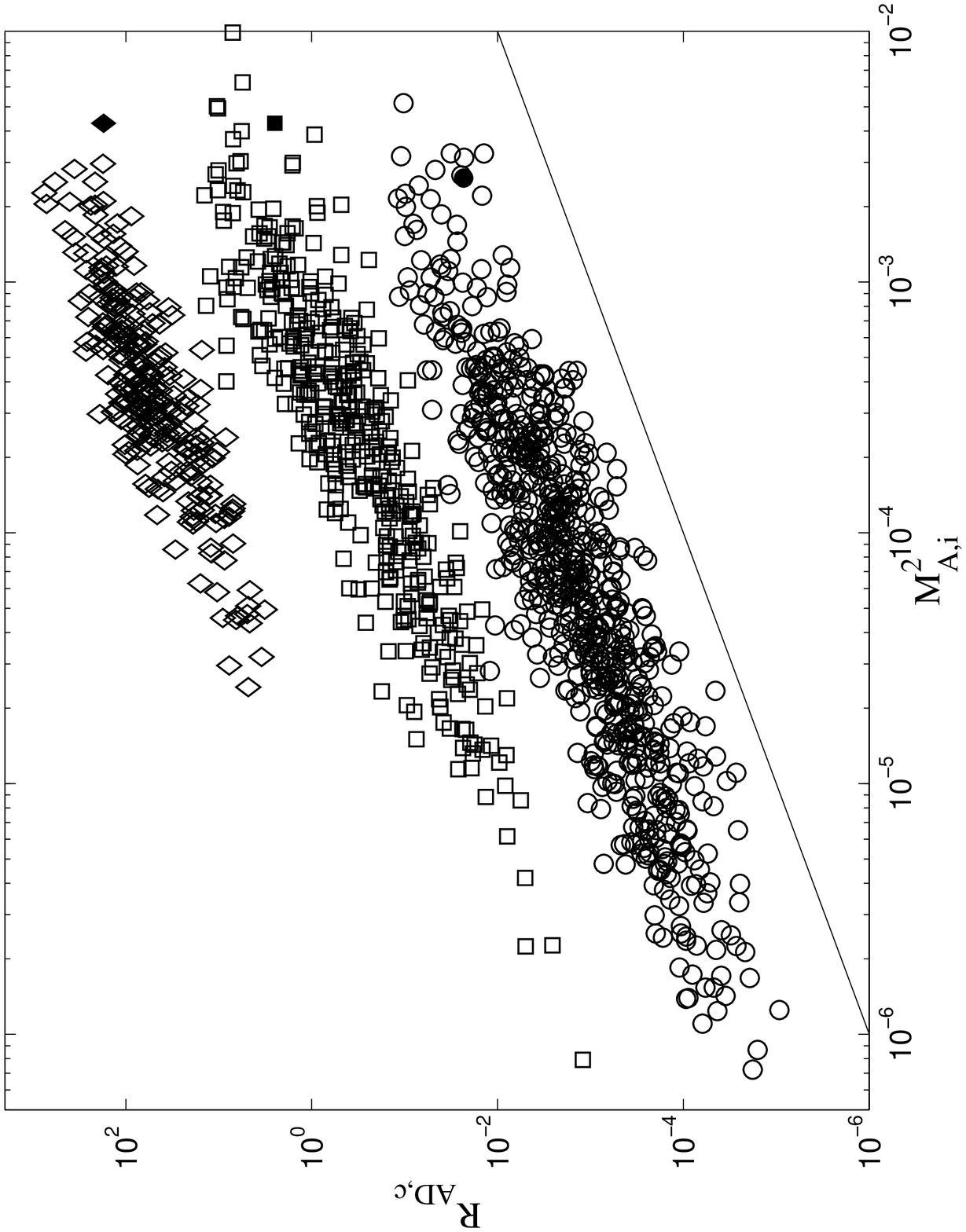}
\caption{Clumps' AD Reynolds number, $\radc$, versus ion \alfven Mach number squared, $\mai^2$, for models m3c2r-1(circles), m3c2r1(squares), and m3c2r3(diamonds) at the end of the simulation.  The straight line shows $\radc$ = $\mai^2$ and
the solid symbols indicate the values of $\rad$ and $\mai^2$ for the whole box.  
\label{fig1}}
\end{figure}

The condition for the validity of
the heavy-ion approximation is $\rad \gg \calm_{\rm Ai}^2$, where
the ion \alfven Mach number, $\mai$, is smaller than the total \alfven
Mach number, $\ma$, by a factor $(\rho_i/\rho)^{1/2}\ll1$ (LMK).
To calculate the AD Reynolds number of a clump, $\radc$, we use the 3D
density-weighted velocity dispersion
of the neutral gas, $\surd 3 \sigma_{n}$, inside a clump as the flow velocity and 
the mean diameter of the clump, $d_c=2r_c$, 
as the length scale.
The ion \alfven Mach number of the clump, $\calm_{\rm Ai,c}$, is taken to be the rms value of $\calm_{\rm Ai}$ of all the cells in the clump.  We can re-write the definition of $\rad$ in equation (\ref{eq:radell}) for clumps as
\beqa
\nonumber\radc(D_c)&\equiv& \frac{4\pi\gad\rho_i\rho_n D_c
\surd 3\sigma_n}{B_{\rm rms}^2} = \frac{\gad\rho_n D_c
\sigma_n}{\surd 3\sigma_i^2}\calm_{\rm Ai,c}^2 \\
&\equiv& C_{\rm HIA}\calm_{\rm Ai,c}^2.
\label{eq:radc}
\eeqa
In Figure \ref{fig1}, we plot $\maic$ versus $\radc$ for models m3c2r-1, m3c2r1, and m3c2r3 at $t = 3\tf$; the results for models m3c2r0 and m3c2r2 lie between the nearby models.
The data points all have $\radc \gg \maic$, even for model m3c2r-1, which has the smallest 
value of $\radlo$. We have verified that this is true at other times as well. 
LMKF found that the Heavy Ion Approximation 
was valid for a turbulent box provided $\rad(\ell_{vi})/\calm_{\rm Ai}^2
\ga 30$, where $\ell_{vi}$ is the length scale for ion-velocity variations, which is generally
significantly smaller than the size of the box. We do not know how $\ell_{vi}$ in the clumps
compares with the clump diameters. If we assume that the two length scales are comparable,
then the requirement for the validity of the Heavy Ion Approximation is $C_{\rm HIA}\ga 30$. This is well satisfied for all the clumps except those in model m3c2r-1, which has $\radl=0.12$ and is the most diffusive run.  For this run, the box as a whole has $C_{\rm HIA}\simeq 10$,
and the Heavy Ion Approximation is at best marginally satisfied. We have not observed any
problems associated with this, however.

Two interesting features of the results are worth noting. First, almost all the clumps have smaller values of $\mai^2$ than the box as a whole; this is expected because of the linewidth-size relation.
The few data points with slightly higher values of $\mai^2$ are due to large statistical fluctuation in the ion density in a few clumps.
Second, we note that the distribution of the data points is roughly parallel to the power law $\rad \propto \mai^2$ (the straight line). This is because the factor $C_{\rm HIA}$ depends on two quantities, the column density, $\rho_n D_c$, and the velocity dispersion ratio, $\sigma_n/\sigma_i^2$, each of which is almost independent of $\mai$.  

\subsection{Clump Mass Function (\clmf)}
\label{sec:cmf}

\subsubsection{Resolution: The Sonic Length and the Inertial Range}
\label{sec:sonic}

	In studying the properties of the clumps that arise in boxes with
supersonic turbulence, two length scales are important:
the sonic length, $\ell_s$, and the minimum scale
for the inertial range, 
$\ell_{\rm in,\, min}$, which corresponds to
the wavenumber $\kinmax=\ell_0/\ell_{\rm in,\, min}$.
The sonic length, which is defined by the condition that the rms turbulent velocity in
a box of size $\ell_s$ equal the sound speed,
gives a characteristic scale for density fluctuations in a supersonically
turbulent medium \citep{pad95,vaz03}. 
The sonic length should be well resolved in numerical simulations since it is important to resolve these density fluctuations and the turbulent motions
that produce them. 
The resolution condition is
$\Delta x\ll \ell_s$, where $\Delta x$ is the size of a grid cell; equivalently,
in terms of the sonic wavenumber $k_s\equiv \ell_0/\ell_s$, we
have $\ell_0/\Delta x\equiv \caln_g\gg k_s$.
We assume that the turbulence in the box exhibits a linewidth-size
relation of the form\footnote{Note that \citet{kru05} defined the sonic length
with respect to the 1D turbulent velocity, $\snt=c_s(\ell_0/\ell_{s,\rm 1D})^q$, 
and adopted $q=\frac 12$;
the two versions of the sonic length are related by $\ell_{s,\rm 1D}=3^{1/(2q)}\ell_s$,
corresponding to $\ell_{s,\rm 1D}=3\ell_s$ for $q=\frac 12$.}
\beq
\calm=3^{1/2}\frac{\snt}{c_s}=\left(\frac{\ell_d}{\ell_s}\right)^q,
\label{eq:machsonic}
\eeq
where $\ell_d$ is the effective minimum driving scale; the corresponding
wavenumber is $k_d\equiv \ell_0/\ell_d$. In our simulations, $k_d=2$, and
we find that the average Mach number in boxes of size $\ell_d$ is indeed very
nearly equal to that for the entire box, $\calm\simeq 3$. We also
find $q\simeq \frac 12$
for $\radlo$ in the range 0.12-12; for $\radlo =120,\;1200$, we find $q\simeq \frac 14$.
The sonic length in a simulation is then
\beq
\ell_s=\frac{\ell_0}{k_d\calm^{1/q}}.
\label{eq:soniclength}
\eeq
Correspondingly, we have
\beq
k_s\equiv \frac{\ell_0}{\ell_s}=k_d\calm^{1/q}\simeq 18-160,
\eeq
for $q=\frac 12$ and $q=\frac 14$, respectively.
This satisfies the resolution condition $\caln_g=512\gg k_s$ for
$\radlo\leq 12$; for $\radlo=120,\;1200$, this resolution condition is
only marginally satisfied.

	Before leaving the topic of the sonic length, we note that it can be inferred
for actual molecular clouds as well.
For $q=\frac 12$ (the observed value--\citealp{hey04}),
the sonic length is related to the linewidth-size parameter $\spc$ (eq. \ref{eq:snt}) by
\beq
\frac{\spc^2}{\mbox{1 pc}}=\frac{2c_s^2}{3\ell_s}, 
\eeq
or
\beq
\ell_s=\frac{2c_s^2}{3\spc^2}~~\mbox{pc}\;=0.0455\left(\frac{T_1}{\spcs^2}\right)~~\mbox{pc}.
\label{eq:lspc}
\eeq

	We define the inertial range of the turbulence as the range of wavenumbers
over which the power spectrum is a power law in $k$. In our simulations, this
extends over the range $\kinmax>k>3$, where $\kinmax\simeq 20$ for our
$512^3$ simulations and $\simeq 10$ for the $256^3$ simulations reported
in LMKF. For $k>\kinmax$, numerical dissipation becomes increasingly important.
Another way of expressing this condition is that with ZEUS, numerical dissipation becomes
important at about 1/10th the minimum wavenumber, $\kinmax\simeq 0.1\times (\caln_g/2)$.
It is desirable to have the sonic length in the inertial range $k_s<\kinmax$,
and this is satisfied for 
the $512^3$ simulations with $\radlo\leq 12$. 
Determining whether this
condition is a general requirement for accurate simulations of supersonic turbulence is
beyond the scope of this paper. We note that this condition becomes
increasingly difficult to satisfy as the Mach number increases.

\begin{figure}
\epsscale{.80}
\plotone{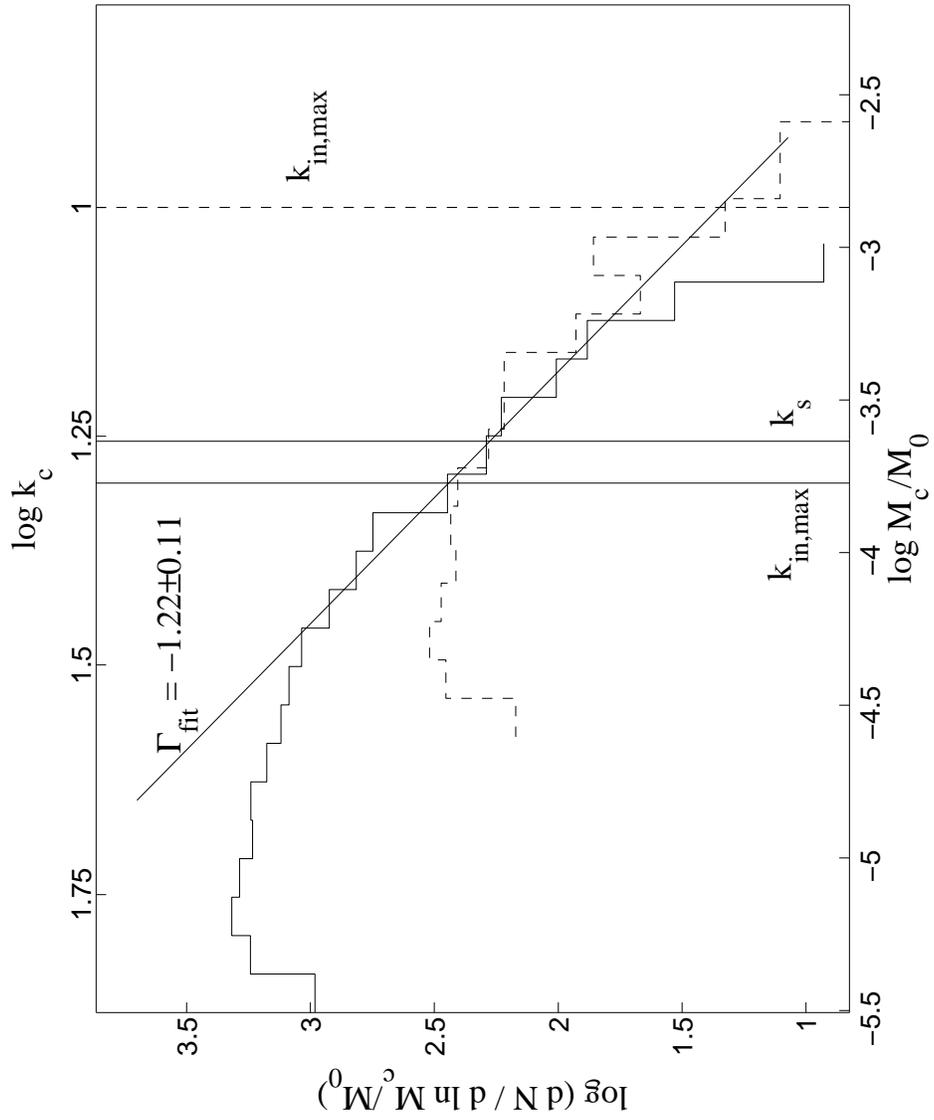}
\caption{Clump mass functions for AD models of $\rad = 1200$ with grid sizes
of $256^3$ (dashed line) and $512^3$ (m3c2r3, solid line).  The sonic wave
number, $k_s$, and the minimum scale of the inertial range, $\kinmax$, for
model m3c2r3 are plotted as vertical solid lines; $\kinmax$ for the $256^3$
model is plotted as a vertical dashed line. The clump wavenumber $k_c$ based
on equation (\ref{eq:mcm0}) is plotted at the top of the figure for
reference; note that $k$ increases to the left. See \S4.2.1 for discussion.\label{fig2}}
\end{figure}

\begin{figure}
\epsscale{1.0}
\plotone{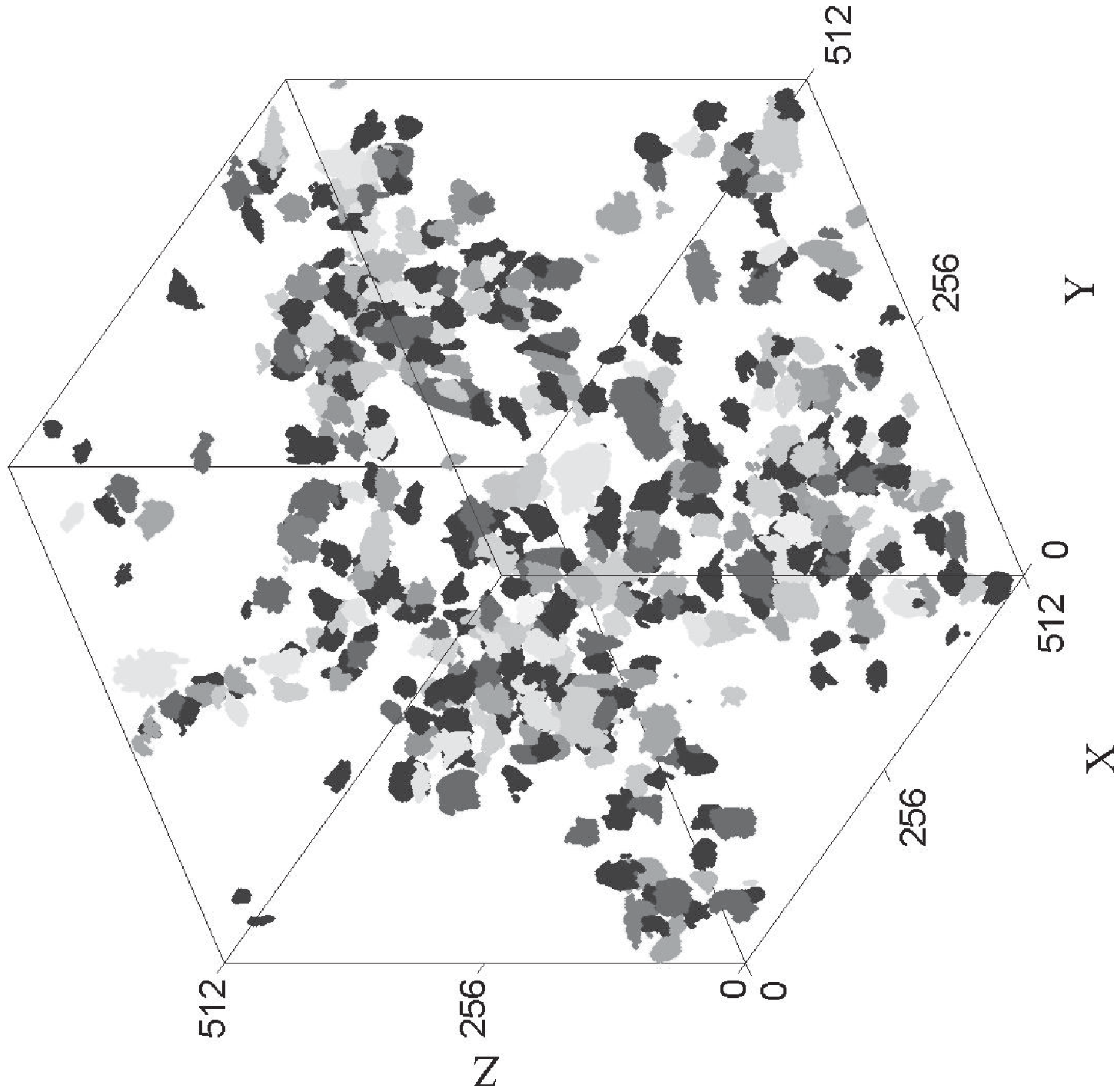}
\caption{3D spatial distribution of clumps, identified by CLUMPFIND with minimum mean radius of 6 cells, from model m3c2r1.  Different gray-scale shadings (different colors in the online version of the paper) are used only to visually separate overlapping individual clumps.\label{fig3}}
\end{figure}

	Figure \ref{fig2} shows the clump mass distribution for the case of
$\radlo=1200$ (close to ideal MHD), at resolutions of $512^3$ and $256^3$.
We can make an approximate relation between the clump masses and wavenumbers
by associating a wavenumber $k_c\equiv \ell_0/D_c$, where $D_c$ is
the clump diameter, to each clump. The corresponding clump
mass is approximately
\beq
\frac{M_c}{M_0}=\frac{4\pi}{3}\left(\frac{\bar\rho_c}{\bar\rho}\right)\frac{1}{(2k_c)^3},
\label{eq:mcm0}
\eeq
where $\bar\rho_c$ is the average clump density. For the high-resolution run, 
the mean density of the clumps within the inertial
range ($k_c<20$) is $\bar\rho_c = 2.6 \bar\rho$
(i.e., the mean density is 2.6 times the minimum clump density).
The higher-mass part of the \clmf\  appears to be a power law (this is justified
in \S \ref{sec:turbfrag} below).
Observe that the slope of
the \clmf\  changes at log $M_c \sim -4.3$, corresponding to $k_c\simeq 30
= 1.5\kinmax$. 
In fact, the clumps with such a mass have $D_c \sim 12 -20$ cells.  
This is similar to both the maximum wavenumber in the inertial range and to the sonic wavenumber, which are also shown in Figure \ref{fig2}, to within a factor of 2. In order to determine whether either of these parameters is associated with the change in slope, we also plot the
clump mass spectrum for the corresponding $256^3$ run, for which
$\kinmax=10$ (vertical dashed line) is reduced by a factor 2 whereas $k_s$ is unchanged.  
The results are clear: The break in the clump mass spectrum in the low-resolution
run occurs at half the wavenumber as in the high-resolution one.
Furthermore, there is no discernable effect associated with the sonic
wavenumber, although it would be desirable to test this conjecture
with both higher resolution simulations and for higher Mach numbers than
$\calm=3$, the value in the present simulations.
It therefore appears that the dominant effect in determining the deviation of
the \clmf from a power law
is the numerical dissipation that sets
in for wavenumbers $k>\kinmax$. The results of our simulations can therefore
address only the higher-mass portion of the \clmf, with a minimum diameter of 12 cells.
Figure \ref{fig3} shows the 3D spatial distribution of clumps, identified by CLUMPFIND with minimum diameter of 12 cells, from a snapshot of model m3c2r1.

\subsubsection{Implications for the Turbulent Fragmentation Model for the IMF}
\label{sec:turbfrag}

As remarked above, the similarity between the core mass function and the stellar IMF suggests that the IMF may be defined during the formation of cores inside molecular clouds.  In the turbulent fragmentation model of \citet{pad02} and \citet{pad07}, 
the mass distribution of cores (i.e., gravitationally unstable clumps)  has the form
\beq
\frac{d\caln_{\rm core}}{d\ln m}\propto\left[1+{\rm erf}\left(\frac{4\ln m+\sigma_x^2}{2\sqrt{2}\sigma_x}\right)\right]m^{-\Gamma},
\label{eq:cmf}
\eeq
where $\sigma_x$ is the dispersion of the density PDF.
The power-law index of the core mass function at high masses, $\Gamma$,
is related to the index, $n_v$, of the velocity power spectrum, 
$P(k)dk\propto k^{-n_v}dk$, by
\beq
\Gamma=\frac{3}{4-n_v}
\label{eq:amhd}
\eeq
for the strong-field, ideal MHD case ($B \geq B_{\rm cr})$, and
\beq
\Gamma=\frac{3}{5-2n_v}
\label{eq:ahd}
\eeq
for the weak-field, ideal MHD case ($B < B_{\rm cr}$), which includes
the hydrodynamic case \citep{pad07}; here
the critical magnetic field $B_{\rm cr}$ is defined by the condition that the postshock gas pressure be comparable to the postshock magnetic pressure.  \citet{hen08} have
introduced an improved theory for the IMF, but we cannot comment on the
differences between their results and those of \citet{pad07} since our simulations
do not include self-gravity. As noted above, the Hennebelle \& Chabrier theory 
predicts a much smaller difference
in the slope of the IMF between the magnetic and non-magnetic cases than does the theory of Padoan et al.

	The simulations of \citet{pad07} do not include self-gravity. Based on
the discussion at the beginning of \S \ref{sec:scaling} below, we note that if one specifies
the temperature and adopts a linewidth-size relation, then it is possible to fix
one parameter arbitrarily. For the Mach number they adopted ($\calm=10$), their box size of
$\ell_0=6$~pc is in good agreement with the linewidth size relation in equation 
(\ref{eq:snt}). However, they
chose a density $\nbh=2\times 10^4$~cm\eee, which results in
a virial parameter $\avir\simeq 0.028$, far lower than observed values.

	In contrast to the Padoan et al simulations, we have included 
an additional physical process---ambipolar diffusion---so that
the strength of the self-gravity is determined by the parameter
governing that process, $\radlo$, as discussed in
\S \ref{sec:adsg}. For the value of $\ma$ we have adopted ($\ma=0.67$)
and for the fiducial values of the ion-neutral coupling
coefficient $\gad$ and the ionization parameter $\chi_i$, the
virial parameter of the box is $\avir=41/\radlo^2$
(eq. \ref{eq:avirtext}). This is unphysically low for $\radlo\ga 6$:
in nature, large values of the AD Reynolds number are
accompanied either by large values of $\ma$ (which is unlikely
according to the results of \citealp{cru99}) or by larger ionizations
than implied by $\chiis=1$.
The core Bonnor-Ebert mass---that is, the
Bonnor-Ebert mass based on the turbulent pressure in the ambient
medium (eq. \ref{eq:mbecore})---in our simulations is
\beqa
\nonumber M_{\rm BE,\, core}&=&\frac{\surd 3}{\calm} \mbe=
0.84\frac{\avir^{1/2}T_1^2}{\spcs^2}~~M_\odot\ \\
&\rightarrow& \frac{5.4}{\radlo}~~M_\odot,
\eeqa
where the last expression is for fiducial values of the parameters.
Of the five AD models in Table 2, only model m3c2r1, with $\radlo=12$
(comparable to the observed values),
yields a physically plausible Bonnor-Ebert mass. We therefore do
not attempt to put our clump mass function in physical units here
(physical units are discussed in \S \ref{sec:scaling} below). What
we can study is the slope of the higher-mass portion of the \clmf, which
is independent of the choice of units.

Ambipolar diffusion could introduce two changes in the value of 
$\Gamma$: First, the relation between $\Gamma$ and $n_v$ could be different, varying from the hydrodynamic relation to the MHD one as $\radlo$ goes from 0 to $\infty$. Second, as shown in LMKF, the value of $n_v$ also depends on
$\radlo$. In the pure hydrodynamic case, $n_v=2$ for supersonic turbulence, whereas in the MHD case the value of $n_v$ is not precisely known and could depend on the plasma $\beta$.

\begin{figure}
\epsscale{.80}
\plotone{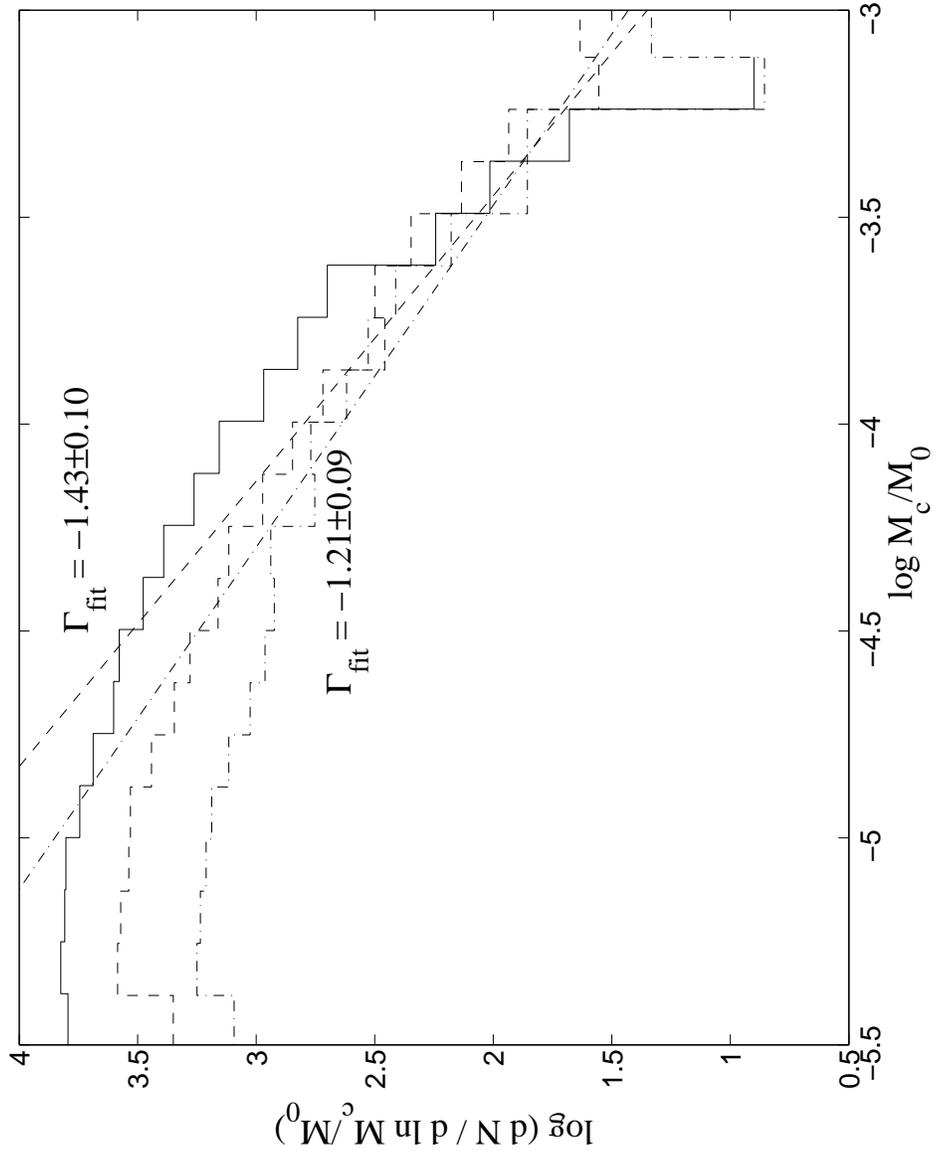}
\caption{Clump mass functions for models m3c2r-1 [$\radlo=0.12$; solid line], 
m3c2r1 [$\radlo=12$], and m3i (ideal MHD).  The dashed line and dot-dashed
line show the best fitting higher-mass slope $\Gamma_{\rm fit}$ for models
m3c2r1 and m3i, respectively.  Model m3c2r-1, which has the strongest ambipolar diffusion,
has the steepest higher-mass slope (see \S\ref{sec:turbfrag} for discussion);
Model m3i is very similar to model m3c2r3 [$\radlo=1200$], which is not shown.
All clumps with radius larger than 3 cells are included in the plot.\label{fig4}}
\end{figure}

In Figure \ref{fig4}, we show the \clmfs\  of models m3c2r-1, m3c2r1, and m3i. LMKF
showed that the density correlation between data sets at different times approaches zero in a time slightly less than $\tf$.  Therefore, in order to build up the statistics, we use three data sets in each model run, at $t\simeq \tf,\; 2\tf$ and $3\tf$.  Adding all the clumps together to form a single data set, we use reduced $\chi^2$ fitting to determine the higher-mass slope of the core mass function, $\Gamma_{\rm fit}$.  
We face two problems in determining $\Gamma_{\rm fit}$: First, we do not know the
the range of masses to include in the ``higher-mass" data, and second, we do not
want our answer to depend on the size of the bins used in binning the data.
We begin by dividing the data 
into 20 logarithmically spaced mass bins.
To address the first problem, we carry out fits beginning with only the three
highest-mass bins, and then steadily increase the number of bins used in the 
fitting until
the peak of \clmf\  is reached.
Initially, the value of $\chi^2$ drops as the number
of bins increases, since more data are contributing to the determination of the slope.
However, when the number of bins is large enough that 
the \clmf\  begins to deviate from a power law, the reduced $\chi^2$ will  increase.
To address the second problem, we increase the number of bins 
from 20 to 40 in increments of 5 and
adopt the value of $\Gamma_{\rm fit}$ with the smallest reduced $\chi^2$ from all five sets of fitting.  Usually, the slopes corresponding to the minimum reduced $\chi^2$ from different total bin numbers are close to each other.  The resulting slopes are listed in Table 3.

In view of the noise fluctuations in the higher-mass range of the \clmf, we have performed a two-sample Kolmogorov-Smirnov (K-S) test to determine whether this part of the \clmf\  can be fit with a power law. The mass range extends from the highest mass bin to the breakpoint determined by the $\chi^2$ fitting procedure described above. The null hypothesis is that the higher-mass end of the \clmf\  from the simulation has the same distribution as a power law. Our results show that the K-S test on all five AD models and the ideal MHD model fails to reject the null hypothesis at the 5\% confidence level. The $p$-values of all the K-S tests with different binning are between 0.49 and 0.97. We conclude that the higher-mass portion of
the \clmf\  is statistically consistent with a power law. 
For the ideal mhd case (model m3i) and for $\radlo=12$ (model m3c2r1), the power
law extends over the entire inertial range. However, in the limit of low $\radlo$
(model m3c2r-1), the power law extends only over the upper half of the inertial
range; higher resolution and/or more samples are needed to determine if the
inertial range is consistent with a power law in this case.

With these $512^3$ models, the clump statistics are adequate to demonstrate that the higher-mass slopes depend on $\rad$.  If turbulent fragmentation is correct, this is no surprise because LMKF 
found that the spectral indexes of the velocity power spectra also depend on $\rad$.
Here we draw on the results of Paper III, which gives more accurate values of
the spectral index for the velocity of the neutrals, $n_{vn}(k)$, than LMKF (see Table 3).  The trend of spectral index changing from an Iroshnikov-Kraichnan \citep{iro63,kra65} to Burgers spectrum \citep{bur74} as one goes from large to small $\rad$ is still clear, as reported in LMKF.

In the limit of ideal MHD (model m3i), the higher-mass slope is $\gfit=1.21\pm0.09$ (see Figure 3), which agrees quite well with
the prediction $\Gamma = 1.18$ from equation (\ref{eq:amhd}) with spectral index $n_v = 1.45\pm0.05$.  Note that \citet{pad07} get somewhat different results ($n_v=1.9$ and $\Gamma=1.4$), but this is presumably due to the difference in flow
conditions: they have $\calm=10$ and $\beta=1$, whereas we have $\calm=3$ and $\beta=0.1$.
In their hydrodynamic simulations, \citet{bal06} found that
the shape of the ClMF depends on the Mach number of the turbulence,
consistent with our result. 
If the shape of the ClMF is significantly affected by the flow conditions, 
then the \citet{pad07} model would imply that the IMF depends on the environment,
since regions of star formation do not all have similar physical 
conditions. As \citet{bal06} point out, this could be problematic in 
view of observational support for an IMF that is approximately universal.

For the model m3c2r1, which has $\rad=12$, comparable to the observed value (\S 2.3), the higher-mass slope is $\gfit=1.43\pm0.10$, which is consistent with the Salpeter value.
As noted above, in the limit of low $\rad$ (model m3c2r-1), 
we are unable to fit the data with a power 
law that extends over the entire inertial range; The slope for the high-mass portion of
the range for which a fit is possible is
$\gfit=2.41\pm0.14$, which continues the trend that the slope increases as $\radlo$
decreases. Since this slope applies to only part of the inertial range,  however, we are
unable to check the validity of equation (\ref{eq:ahd}), which relates
the slope of the ClMF to the velocity power spectrum in the weak field case.
Consistent with the results of
\citet{pad07}, this slope is significantly
greater than the Salpeter value of the higher-mass slope, $\Gamma= 1.35$.

Comparison of numerical simulations of turbulence with either pure hydrodynamics or ideal MHD has 
shown that magnetic fields suppress fragmentation \citep[e.g.][]{pas95,gam03,pad07,hent08}.  We can see this effect in our simulations by comparing the \clmfs\  of models m3c2r-1, in which the neutrals are almost purely hydrodynamic, and m3i, with ideal MHD.  Figure \ref{fig4} shows that the number of clumps in the low-$\rad$ models is greater than in the high-$\rad$ models, except at the higher-mass end:  The total number of clumps with $D_c > 12$ cells from the three time snapshots in the quasi-hydrodynamic model m3c2r-1 is 2093 (Table 3), whereas it is 1033 in model m3i.
The total mass of clumps in model m3c2r-1 is $\sim0.106 M_0$, whereas it is $\sim0.092 M_0$ in model m3i.  On average, the mass per clump in the quasi-hydrodynamic model m3c2r-1 is smaller than that in the ideal MHD model, which is also consistent with prior simulations.  

The turbulent fragmentation model for the IMF predicts that the core mass function (CMF)
(i.e., the mass function of gravitationally bound clumps) is the same
as the clump mass function (\clmf) at high masses, and it is based on the assumption
that the IMF is proportional to the core mass function (the latter is predicted in
the work of \citealp{mat00}). \citet{pad07} emphasize that the predicted \clmf\ for
hydrodynamic turbulence is much steeper than for ideal MHD turbulence,
and our work confirms this. Our work shows that there is a continuous variation
in the higher-mass slope of the \clmf\ due to the effects of ambipolar diffusion, such
that the fraction of stars born at high mass should increase
with $\rad$.

\subsection{Mass-To-Flux Ratios}

Ambipolar diffusion plays an important role in the core collapse process when 
the clump mass is less than or comparable to the magnetic critical mass (eq. \ref{eq:mphi}). 
Observationally, only a limited number of cores 
have measured mass-to-flux ratios due to the difficulty in
making precise Zeeman measurements. Furthermore, 
observations give only  the line-of-sight values
for the magnetic field and column density, so the value of $\mu_\Phi\propto M/\Phi$ 
for any particular core is necessarily uncertain. 
From a study of 34 dark cloud cores, \citet{tro08} found an average value of $\mu_{\Phi,c}=
1.4-2.1$ after allowance for projection effects; the smaller value is based on flattened clouds,
whereas the larger one is for spherical ones. The median values are larger by about 20\%.
Observed cores are thus somewhat magnetically supercritical.

\subsubsection{Resolution Study}

\begin{figure}
\epsscale{.80}
\plotone{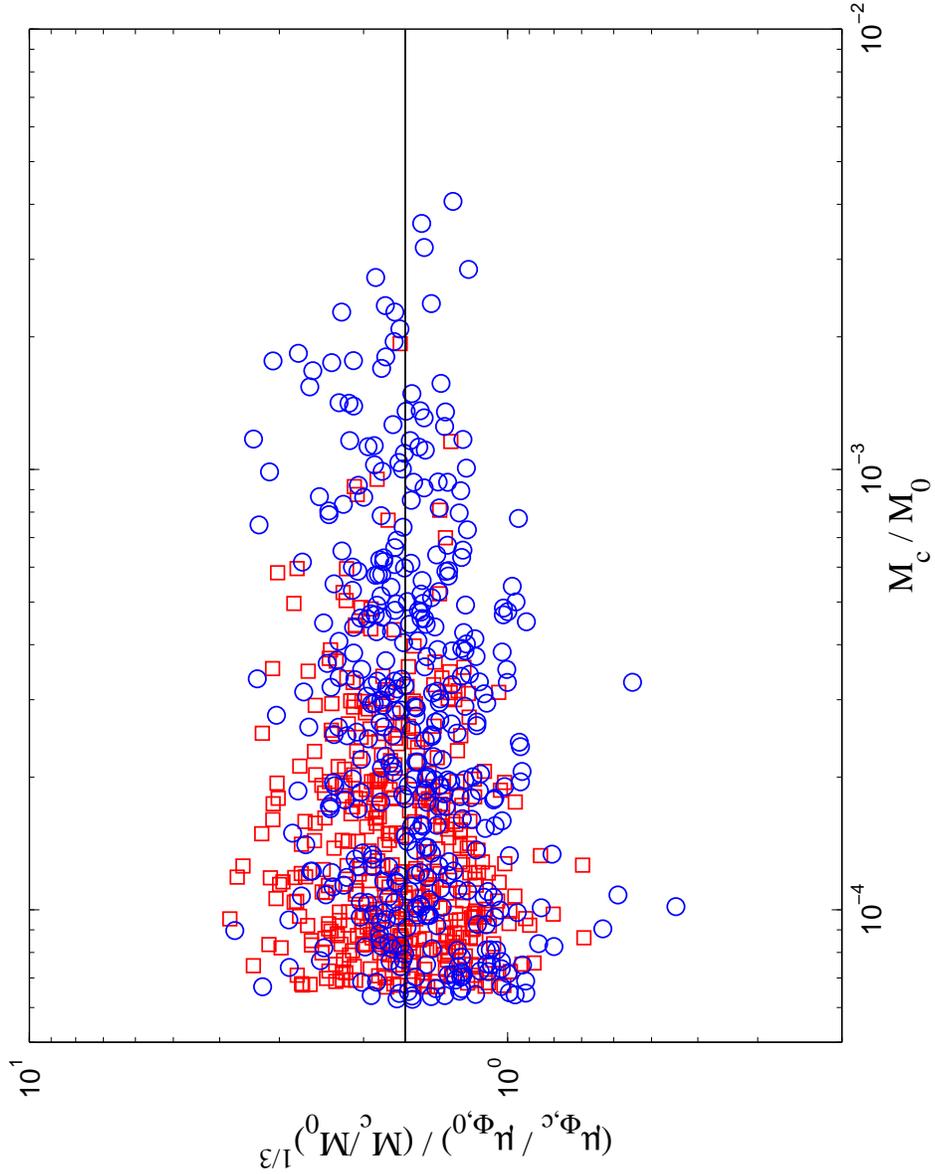}
\caption{Convergence study for the mass-to-flux ratio of the clumps.  Values of normalized mass-to-flux ratios of clumps in the $\rad = 1200$ model are plotted against normalized clump mass, for resolutions of $256^3$ (blue cirlces) and $512^3$ (red squares; model m3c2r3).  By plotting (\mfr)/$(M_c/M_0)^{1/3}$ versus $M_c/M_0$, data points are projected horizontally for easy visual comparison.  The straight line shows the mean of the normalized mass-to-flux ratio for elliptical clumps using the mean density and B-field of the clumps from model m3c2r3.  See $\S4.3.1$ for discussion.
\label{fig5}}
\end{figure}

In this section, we check the convergence of the mass-to-flux ratios of the clumps by comparing the results from the $256^3$ and $512^3$ runs for model m3c2r3.  
To carry out the resolution study, we consider only clumps that have a mass at 
least equal to the minimum mass of clumps with $r_c\geq 12$~cells in the $512^3$ model; for the $256^3$ run, this corresponds to $r_c\ga 6$ cells. 
For easy comparison of clump mass-to-flux ratios among models, we 
eliminate the dependence of the mass-to-flux ratio on clump mass by
plotting the ratio (\mfr)/$(M_c/M_0)^{1/3}$ versus $M_c/M_0$ for both the $256^3$ and $512^3$ runs in Figure \ref{fig5}.  
Curve fitting shows that the slope of the $256^3$ data is $0.06\pm0.03$ and the slope of 
$512^3$ data is $0.07\pm0.02$.  
The mean values of (\mfr)/$(M_c/M_0)^{1/3}$ are $1.68\pm0.03$ and 
$1.78\pm0.03$ for the $256^3$ and $512^3$ models, respectively.  We conclude that the mass-to-flux ratios of clumps in the $512^3$ model are converged.

\subsubsection{Effect of $\radlo$ on the Mass-to-Flux Ratio}
\label{sec:effect}

As discussed in \S \ref{sec:turbfrag} above, our choice of parameters allows
us to study the effect of varying $\radlo$ on the mass-to-flux ratios, but at the expense of
considering models that would be unphysical were gravity to be included:
Equation (\ref{eq:muphisim}) implies $\mu_{\Phi,\,0}=0.17\radlo/(\gad^*\chiis)$
for our simulations, which is in the observed range only for the $\radlo=12$ case.
What is of interest then is how the normalized values of the mass-to-flux ratio 
vary with $\rad$. For example, the ratio of the mass-to-flux ratio for an individual clump,
$\mu_{\Phi,c}$, to that for the entire box is
\beq
\frac{\mu_{\Phi,c}}{\mu_{\Phi,\,0}}=\frac{M_c}{B_c \pi R_{c,\perp}^2}\cdot
\frac{B_0 \ell_0^2}{M_0} = \left(\frac{B_0}{B_c}\right)\frac{\Sigma_c}{\Sigma_0}\simeq
\frac{\Sigma_c}{\Sigma_0},
\label{eq:muphic}
\eeq
where $R_{c,\perp}$ is the radius of the clump normal to the field threading the clump, $B_c$, and $\Sigma_0$ is the mean surface density for the turbulent box. 
For the cases we consider, the mean field in the clump is close to the mean field of the whole box
since the relatively small value of the \alfven Mach number, $\ma=0.67$, leads to 
a relatively uniform field, as discussed in \S \ref{sec:otherp} below.
As a result, for most clumps the mass-to-flux ratios are just proportional to the surface densities.  The ratio $\Sigma_0/\Sigma_c$ is just the number of clumps along a flux tube.
Furthermore, since the density of the clumps is typically a few times the threshold
density (see below eq. \ref{eq:mcm0})
and is thus approximately constant, it follows that the mass-to-flux
ratio in the clumps is proportional to the cube root of the clump mass:
\beq
\mu_{\phi,,c}\propto\Sigma_c\propto \rho_c R_c
\propto R_c\propto M_c^{1/3}.
\eeq

We now use our simulations to determine whether ambipolar diffusion 
in a turbulent medium affects the mass-to-flux ratio in clumps, even in the absence of
self-gravity.  
In order to ensure that the clumps we study are in the higher-mass, power-law regime
of the clump mass distribution so that numerical effects are minimal, we choose
a minimum clump mass that is above the threshold for the higher-mass regime in
all cases. This minimum mass corresponds to a minimum clump radius of
$r_c = 6$ cells. 

\begin{figure}
\epsscale{.80}
\plotone{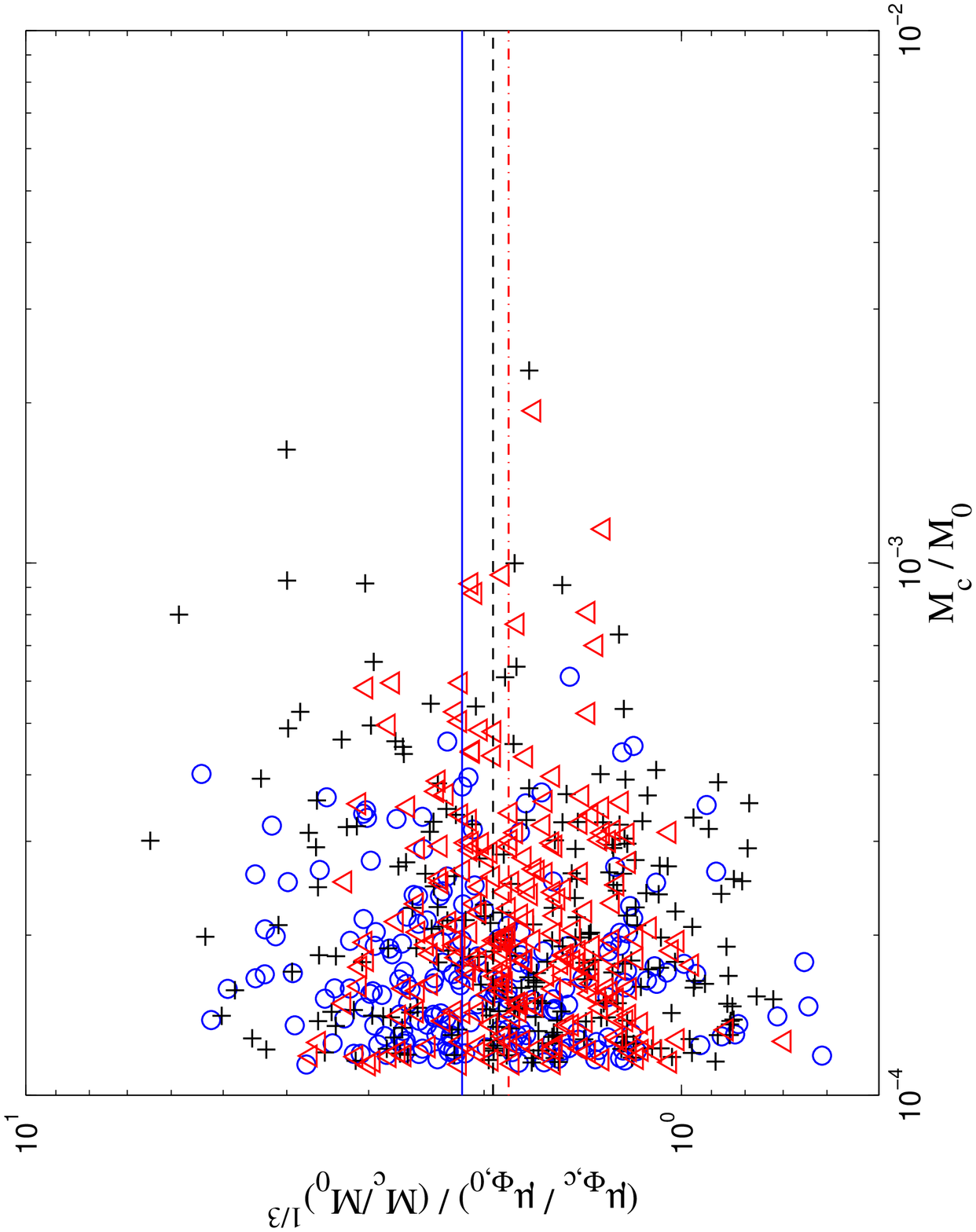}
\caption{Normalized mass-to-flux ratios, (\mfr)/$(M_c/M_0)^{1/3}$, of clumps in models m3c2r-1 (blue circles), m3c2r1 (black crosses), and m3c2r3 (red triangles) plotted versus normalized clump mass $M_c/M_0$.  The mean values of $\avg{\mu_{\Phi,c} / \mu_{\Phi, 0}/(M_c/M_0)^{1/3}}$ for the three models are plotted as the horizontal lines ($\radlo = 0.12$ blue solid, $\radlo = 12$ black dashed, and $\radlo = 1200$ red dot-dashed).  
The model with the largest value of $\rad$ has a slightly lower average mass-to-flux ratio and a smaller dispersion of the mass-to-flux ratios.
\label{fig6}}
\end{figure}

To determine how ambipolar diffusion affects the mass-to-flux ratio,
we compute the value of (\mfr)/$(M_c/M_0)^{1/3}$ for all clumps in each model
and plot the results in Figure \ref{fig6}.
The mean values of \mfr\ for the three models
are also tabulated in Table 3 and are shown as the horizontal lines in Figure \ref{fig6}.
In all the models, the values of $\mu_{\Phi,c}$ for the clumps are smaller than $\mu_{\Phi,0}$ for the whole box due to fragmentation along flux tubes. This effect has been observed in other MHD turbulence simulations \citep{vs05a,till07}.
The typical value of the normalized mass-to-flux ratio, $\mu_{\Phi,c}/\mu_{\Phi,0}\sim 0.1$,
is set by our resolution, since the number of clumps increases with decreasing size and
$\mu_{\Phi,c}$ scales as $M_c^{1/3}$.
We observe from Figure \ref{fig6} and Table 3 that $\avg{\mu_{\Phi,c}} / \mu_{\Phi, 0}$ shows a
small systematic increase from the 
large $\rad$ model to the small and moderate $\rad$ models.  

This table also shows that the mean density of clumps, $\avg{\rho_c}$, in the three models increases systematically as $\rad$ decreases.  
The dispersion in the values of \mfr$/(M_c/M_0)^{1/3}$ in Figure \ref{fig6} shows a significant variation as $\rad$ decreases.  The dispersions of mass-to-flux ratio (not mass-to-flux divided by $M^{1/3}$) are given in Table 3. The dispersion for $\radlo=12$ is almost twice that for $\radlo=1200$.
The larger dispersion of $\mu_{\Phi,c}$ and higher density of clumps 
at $\radlo=12$ than at high $\radlo$ suggest that material can more easily cross magnetic field lines as $\rad$ decreases. A further decrease in $\radlo$ to 0.12 results in a higher
density, but increased fragmentation of the clumps reduces the dispersion somewhat.
We conclude that, even in the absence of self-gravity, ambipolar
diffusion has an effect on the mass-to-flux ratios of clumps.

\subsection{Other Physical Properties of Clumps}
\label{sec:otherp}
\begin{figure}
\epsscale{.80}
\plotone{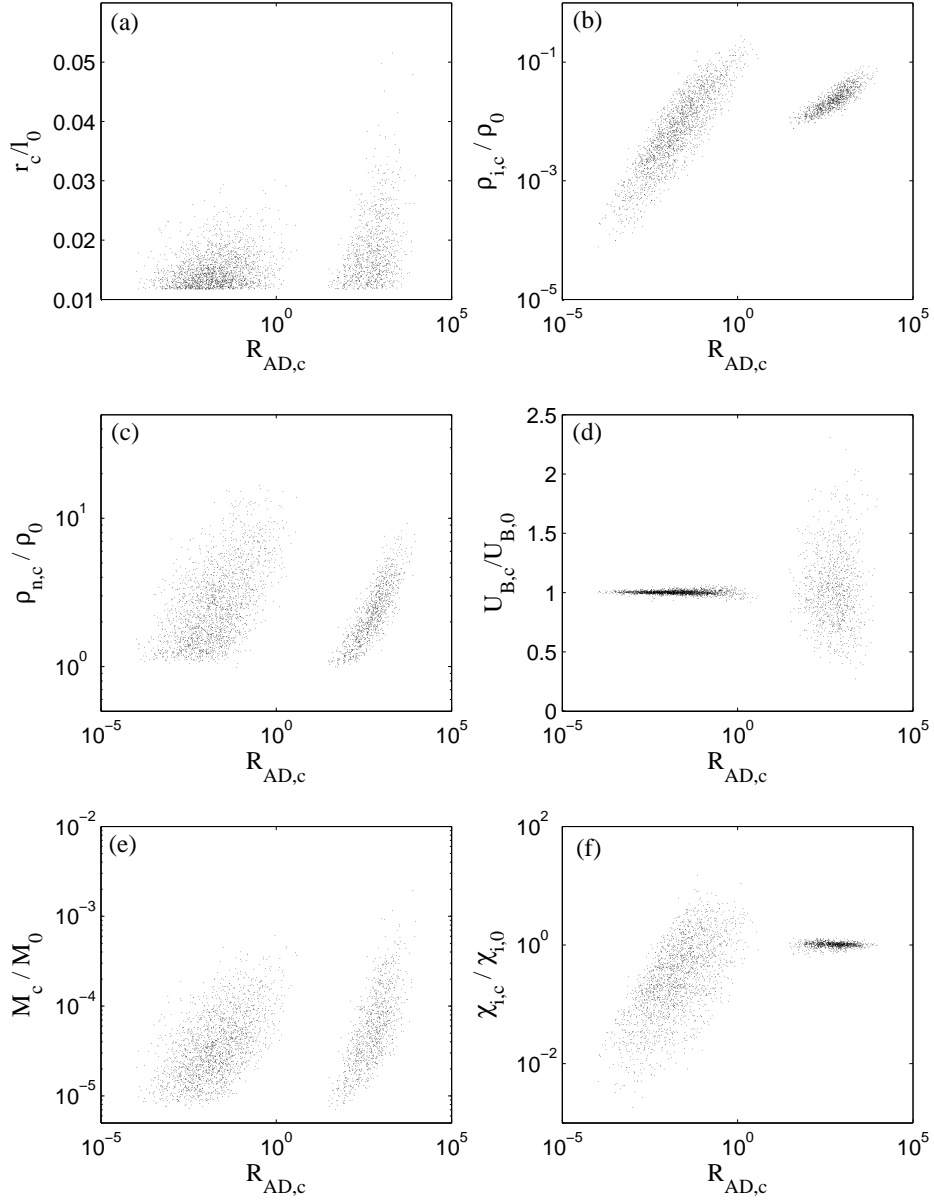}
\caption{Other normalized physical properties of clumps as functions of the AD Reynolds number of the clumps, $\radc$, for the models m3c2r-1 (on the left) and m3c2r3 (on the right): (a) radius, $r_c/\ell_0$, (b) ion density; $\avg{\rho_i}_c/\bar\rho$; (c) neutral density, $\avg{\rho_n}_c/\bar\rho$; (d) magnetic energy density, $U_{B,c}/U_{B,0}$; (e) clump mass, $M_c/M_0$; and (f) ionization mass fraction, $\chi_{i,c}/\chi_{i,0}$.  See \S\ref{sec:otherp} for discussion.
\label{fig7}}
\end{figure}

In this section, we summarize a number of other physical properties of the clumps as functions of $\radc$ in Figure \ref{fig7} by comparing the two models m3c2r-1 [$\radlo=0.12$, strong AD] and m3c2r3 [$\radlo=1200$, strong ion-neutral coupling], which represent the two extremes of ion-neutral coupling among our simulations.  Figure \ref{fig7} gives side-by-side plots of the normalized clump radii, 
$r_c/\ell_0$, the ion and neutral densities, $\rho_{i,c}/\rho_0$ and $\rho_{n,c}/\rho_0$, 
magnetic energy density, $U_{B,c}/U_{B,0}$, clump mass, $M_c/M_0$, and ionization mass fraction, $\chi_i$, for
clumps for the two models.

Figure \ref{fig7}a shows that the normalized radii of the clumps in the strong-coupling model (m3c2r3) are, on average, larger than those for the strong AD model (m3c2r-1).  This is a result of more fragmentation in the strong AD case.  The largest radius in m3c2r3 is about double that in m3c2r-1.  Since we require clumps to have a radius larger than 6 cells, there is a sharp truncation in the 
size distributions at $r_c/\ell_0=6/512=0.012$.  
\citet{vs05b} found that clumps in the non-magnetic case were smaller than those
in the ideal MHD case, consistent with our result.

Note that the clumps in model m3c2r3 have a smaller range of $\radc$ because of the strong coupling between ions and neutrals.  This is seen in all other properties as well.  Figures \ref{fig7}b and \ref{fig7}c show the normalized mean ion and neutral densities of the clumps.  The sharp bottom edge in 
Figures \ref{fig7}a and \ref{fig7}c is
 the result of the density threshold $\rho_c\geq\rho_0$ we chose in defining the clumps. The variations in ion density are much smaller in the strong coupling case than in the strong AD case.  This is also reflected in the ionization mass fraction in Figure \ref{fig7}f.  The ionization mass fraction of clumps in the strong coupling model is about constant, but that of clumps in the strong AD model varies by almost 3 orders of magnitude.
In Figure \ref{fig7}d, the magnetic field is barely perturbed by the turbulence in model m3c2r-1 because of weak coupling; the magnetic field energy density in the clumps, $U_{B,c}$, is very nearly the same as that for the whole box.  Although the magnetic field is perturbed more in model m3c2r3, most clumps have $U_{B,c}$ within 50\% of that in the box.
Figure \ref{fig7}e shows that the largest clumps in m3c2r3 are more massive than the largest ones in m3c2r-1.
This larger mass is due to a larger size, since the densities in the two models are about the
same, and can be understood as
the result of magnetic suppression of fragmentation, as discussed in \S 5.2.2.  The clump 
properties shown in Figure \ref{fig7} include clumps down to $r_c = 6$ cells.  From this figure, we see that the global physical properties of clumps scale smoothly from $r_c = 6$ 
cells to the largest clump.

\section{Physical Units for Simulations of Turbulent Boxes with Ambipolar Diffusion}
\label{sec:scaling}

The results of our simulations have been reported in dimensionless form.
How can they be converted to physical values?
A simulation of an isothermal, magnetized, turbulent box is characterized by three dimensional parameters---the size of the box, $\ell_0$, the mean density in the box, $\nbh$, and the isothermal sound speed, $c_s$---and two dimensionless ones---the 3D sonic Mach number, $\calm=3^{1/2}\snt/c_s$ and the plasma-$\beta$ parameter, $\beta\equiv 8\pi\bar\rho c_s^2/\brms^2$ (\citealp{ost99,pad99}).
Here $\nbh$ is the mean density of hydrogen nuclei, $\snt$ is the 1D nonthermal velocity dispersion and $\brms\equiv\avg{B^2}^{1/2}$ is the rms magnetic field. 
 In the absence of other physical processes, all these parameters can be selected arbitrarily, although  the value of
$c_s\propto T^{1/2}$ is tightly constrained for molecular clouds,
which generally have temperatures in the range $10-20$~K. 

Inclusion of a new physical process, such as ambipolar diffusion, introduces a new dimensional constant,
in this case the ion-neutral coupling parameter, $\gad$. Correspondingly, a new dimensionless
parameter (in this case, $\rad$) can be formed and the number of independent dimensional
parameters is reduced by one. 
For a given sound speed, there is thus one independent dimensional parameter,
such as the density, in simulations of ambipolar diffusion; such simulations
are therefore scale free. Treatments of ambipolar diffusion require 
specification of the ionization, which
in principle can introduce another dimensionless parameter that in turn
would determine the scale. However, as discussed in \S \ref{sec:codeunits}, the Heavy Ion
Approximation eliminates this constraint.

Adoption of a linewidth-size relation, as is observed in
molecular clouds \citep{lar81}, also reduces the number of independent dimensional
parameters by one. Hence, if an isothermal system satisfies a linewidth-size relation and is subject to ambipolar diffusion, then its velocity scale is set by
the isothermal sound speed, $c_s\propto T^{1/2}$, and its size and mean density are
determined by dimensionless parameters. In this case, a given simulation applies
to only one set of parameters describing the box. This is discussed further in
the Appendix, which gives explicit expressions for properties of turbulent boxes
in the general case, when they satisfy a linewidth-size relation, and for
self-gravitating boxes. Here we present the scaling for our simulations
of turbulent boxes
with ambipolar diffusion.

\subsection{General Scaling Relations}
\label{sec:adscale}
	To determine how simulations of a turbulent box with ambipolar diffusion
can be scaled to physical systems, we 
use equation (\ref{eq:radladtext}) to solve for the size of the simulation box, $\ell_0$. We find that it is determined by the
remaining two dimensional parameters ($\nbh$ and $T$) along with
five dimensionless parameters [$\radlo$, $\gad^*$, $\chiis$, $\beta$ and $\calm$]:
\beq
\ell_0=0.031\left[\frac{\radlo}{\gad^*\chiis}\right]\frac{\calm}{\ma^2}\left(\frac{T_1}
{\nbht}\right)^{1/2}~~\mbox{pc},
\label{eq:load}
\eeq
The flow time across the box, the mass in the box, and the
column density are then
\begin{eqnarray}
t_f&=&\frac{\ell_0}{\calm c_s}=1.62\times 10^5\left[\frac{\radlo}{\gad^*\chiis}\right]
\frac{1}{\ma^2\nbht^{1/2}}~~~\mbox{yr},\\
 M_0&=&\bar\rho \ell_0^3~=~1.06\times 10^{-3}\left[\frac{\radlo}{\gad^*\chiis}\right]^3
\frac{\calm^3T_1^{3/2}}{\ma^6\nbht^{1/2}}~~M_\odot,
\label{eq:mad}\\
\Nh&=&\nbh \ell_0=9.6\times 10^{19}\left[\frac{\radlo}{\gad^*\chiis}\right]
\frac{\calm}{\ma^2}(\nbht T_1)^{1/2}  
~~~\mbox{cm\ee},
\label{eq:Nad}
\end{eqnarray}
Note that these scalings are
preserved by the Heavy-Ion Approximation, in which the
ion mass fraction ($\propto\chiis$) is increased and the ion-neutral
coupling coefficient ($\propto\gad^*$) is decreased by the same factor.
The strength of the magnetic field does not depend on $\radlo$,
\beq
B= (4\pi\rho c_s^2)^{1/2}\frac{\calm}{\ma}
=3.2(\nbht T_1)^{1/2}\frac{\calm}{\ma}~~~\mug.
\label{eq:badt}
\eeq

As discussed in \S \ref{sec:adsg}, simulations of gas in which the ionization
scales as $\nh^{-1/2}$ have an implicit value of the virial parameter, $\avir$,
given by equation (\ref{eq:avirtext}). 
Actual physical systems have $\avir\ga 1$, since violations of this inequality
lead to gravitational motions that raise $\avir$ up to order unity.
Hence, this sets a lower limit on the product of the ionization parameter and the
AD coupling parameter for a given value of $\radlo$,
\beq
\gad^*\chiis\ga \frac{\radlo}{14.2\ma^2}.
\label{eq:gadchi}
\eeq
The lower limit on the ionization corresponds to the case of gravitationally
bound clouds and clumps discussed in \S \ref{sec:adsg} (for
spherical clouds, the coefficient 14.2 is replaced by $14.2/c_V^{1/2}\simeq 19.7$).

\subsection{Scaling with the Linewidth-Size Relation}

	Most molecular gas in the Galaxy is observed to obey a linewidth-size relation
\beq
\snt=\spc R_{\rm pc}^{1/2},
\label{eq:snt}
\eeq
where $R_{\rm pc}$ is the radius of the region measured in pc and
typically $\spc\simeq 0.72$~km~s\e\ \citep{mck07}.
The linewidth-size relation is quite general: it 
applies to within a factor $\sim 3$ to molecular gas ranging
from small clumps much less than 1 pc in size to GMCs \citep{fal09}.
Taking $R=\ell_0/2$
and noting
that $\calm$ is the 3D Mach number, we find
\beq
\calm=3^{1/2}\;\frac{\snt}{c_s}=\left(\frac{3\ell_0}{2c_s^2}\right)^{1/2}\frac{\spc}{(1\;\rm pc)^{1/2}}=
4.69\;\frac{\spcs \lpc^{1/2}}{T_1^{1/2}},
\label{eq:calm}
\eeq
where
\beq
\spcs\equiv\frac{\spc}{0.72~\mbox{km s\e}}\;.
\label{eq:spcs}
\eeq
\citet{fal10} have shown that
this {\it turbulence-dominated linewidth-size relation}
applies only when
\beq
\Nh<N_\lws=1.3\times 10^{22}\spcs^2~~~\mbox{cm\ee},
\label{eq:Nlwst}
\eeq
or, equivalently, when
\beq
\nbh<\bar n_\lws=9.6\times 10^4\left(\frac{\spcs^4}{\calm^2 T_1}\right).
\label{eq:nlwst}
\eeq
For larger values of the column density and density, the linewidth-size 
relation must take the effects of self-gravity into account. The resulting
{\it virialized linewidth-size relation} has $\sigma\propto (\Sigma\ell)^{1/2}$ and
is equivalent to setting the virial parameter equal to unity, $\avir=1$ 
(\citealp{hey09}; see \S \ref{sec:lwsrelations}).
The linewidth is greater than that in the turbulence-dominated case due to 
the effects of self gravity.

When the turbulence-dominated linewidth-size relation applies,
so that $\Nh$ and $\nbh$ satisfy the inequalities in
equations (\ref{eq:Nlwst}) and (\ref{eq:nlwst}), then 
the size of the simulation box is determined by equation (\ref{eq:calm}):
\beq
\ell_0=0.0454\left(\frac{\calm^2 T_1}{\spcs^2}\right)~~~\mbox{pc}.
\label{eq:lolws}
\eeq
With the aid of equation (\ref{eq:load}), one can then
express the density in terms of the linewidth-size parameter, $\spcs$,
\beqa
\nbh&=&9.6\times 10^4
\left(\frac{\spcs^4}{\calm^2 T_1}\right)\left[\frac{\radlo}{14.2\ma^2\gad^{*}\chiis}\right]^2
~~~\mbox{cm\eee},\\
&=&\bar n_\lws\left[\frac{\radlo}{14.2\ma^2\gad^{*}\chiis}\right]^2,
\label{eq:nbhscale}
\eeqa
where the factor in brackets is $\la 1$ since the corresponding virial parameter
must be $\ga 1$ (eq. \ref{eq:gadchi}).  Similarly, one can show that
\beq
\Nh=N_\lws\left[\frac{\radlo}{14.2\ma^2\gad^{*}\chiis}\right]^2
\eeq
with the aid of equation (\ref{eq:Nlwst}).
The mass corresponding to $N_\lws$ and $\bar n_\lws$---i.e.,
the maximum mass at which the turbulence-dominated linewidth size
relation holds---is $M_\lws$, which is given in equation (\ref{eq:mlws}).
The mass in the simulation box is given in terms of $M_\lws$ by
\beq
M_0=M_\lws\left[\frac{\radlo}{14.2\ma^2\gad^{*}\chiis}\right]^2.
\eeq

On the other hand, when the system being simulated is self-gravitating, then
$\avir\sim 1$ and the inequality in equation (\ref{eq:gadchi}) is replaced by
an equality. Equations (\ref{eq:load}), (\ref{eq:mad}), and  (\ref{eq:Nad})
show that in this case, $\ell_0\propto \nbh^{-1/2}$ and $M_0\propto \nbh^{-1/2}$
are smaller than in the turbulence-dominated case, whereas $\Nh\propto \nbh^{1/2}$
is larger. The general case is discussed in the Appendix, \S\ref{sec:scalevlws}.

\subsection{Physical Parameters for Simulations}

We are  now in a position to discuss the physical parameters corresponding
to our simulations. 
For simplicity, we shall assume that 
the temperature is $T=10$~K and that the linewidth-size parameter
has its fiducial value, $\spcs=1$, corresponding to $\spc=0.72$~km~s\e.
The maximum column density for the turbulence-dominated linewidth-size
relation is $N_\lws=1.3\times 10^{22}$~cm\ee, and 
since the Mach number is $\calm=3$,
the corresponding 
maximum density
is $n_\lws= 1.1\times 10^4$~cm\eee.
We shall focus on the four cases $\radlo=1.2,\,12,\,120,\,1200$,
since the $\radlo=0.12$ case was done to model the transition to the 
hydrodynamic limit. 
Recall that the clouds in \citet{cru99}'s observations have
a logarithmic mean value $\avg{\rad}_{\log}=17$, comparable to 
the value in the $\radlo=12$ simulation. The $\radlo=1.2$ simulation
has a somewhat smaller value of $\rad$, and the $\radlo=120$ simulation
a somewhat larger value, than any of the clouds in that sample;
however, it must be borne in mind that this sample by no means covers
all the types of molecular gas in the Galaxy. 
In particular, the $\radlo=1200$ run is relevant to the outer parts of GMCs,
where the ionization is dominated by C$^+$.

The virial parameter associated with a given value of $\radlo$ in
our simulations is
\beq
\avir=\left[\frac{\radlo}{6.4\gad^*\chiis}\right]^{-2}
\label{eq:avirscale}
\eeq
from equation (\ref{eq:avirtext}).
Since our simulations have $\ma=0.67$, the constraint on the ionization set by the requirement
 $\avir \ga 1$ implies (eq. \ref{eq:gadchi})
\beq
\radlo\la 6.4\gad^*\chiis.
\label{eq:radlomax}
\eeq

First consider the case in which $\radlo=1.2$. We assume
that the ionization and coupling parameters have their fiducial values
($\gad^*=\chiis=1$); the ionization constraint is then well satisfied.
The virial parameter is $\avir=28$ from equation (\ref{eq:avirscale}),
so the self-gravity is negligible in the system being simulated.
Equations (\ref{eq:load}), (\ref{eq:mad}), (\ref{eq:Nad}) and (\ref{eq:badt}) imply
that the  size of the system is $\ell_0=0.25\nbht^{-1/2}$~pc,
the mass is $M=0.54\nbht^{-1/2}\, M_\odot$, the
column density is $\Nh=7.7\times 10^{20}\nbht^{1/2}$~cm\ee,
and the magnetic field is $B=14\nbht^{1/2}\;\mu$G.
Much of the unbound molecular gas in the Galaxy satisfies
the turbulence-dominated linewidth-size relation \citep{fal09}.
If the simulated system satisfies this relation, 
then the density is $\nbh=370$~cm\eee\ from equation
(\ref{eq:nbhscale}), and correspondingly
the size of the simulation box is $\ell_0=0.4$~pc, 
the mass is $M_0=0.9 \;M_\odot$, 
the column is $\Nh=4.7\times 10^{20}$~cm\ee, and
the magnetic field is $B=8.7\;\mu$G.

Next, consider the simulations with $\radlo=12,\,120$. For these
runs, the implied virial parameter is less than unity
for the fiducial values of the ionization and coupling parameters.
If $\gad^*\chiis$ is as close as possible to its fiducial value, then
the inequality in equation (\ref{eq:radlomax}) becomes an equality,
and equations (\ref{eq:load}), (\ref{eq:mad}), and (\ref{eq:Nad}) imply
$\ell_0=1.3\nbht^{-1/2}$~pc, $M_0=82\nbht^{-1/2}\, M_\odot$,
and $\Nh=4.1\times 10^{21}\nbht^{1/2}$~cm\ee. These conditions correspond
to $\avir=1$, so the systems are on the virialized linewidth-size relation.
As remarked above, the virialized linewidth-size relation applies
when the density and column density are large,
$\nbh\ga\bar n_\lws=1.1\times 10^4$~cm\eee 
and $\Nh\ga N_\lws=1.3\times 10^{22}$~cm\ee. Correspondingly,
the size of the system is $\ell_0\la 0.4$~pc
the mass is $M_0\la M_\lws=25 M_\odot$,
and the magnetic field is $B\ga 50\;\mu$G.

In sum, the systems we have simulated are relatively small, 
with $\ell_0\la 1/\nbht^{1/2}$~pc (eqs. \ref{eq:load} and \ref{eq:gadchi})
and $M_0\la 100/\nbht^{1/2}\;M_\odot$ (eq. \ref{eq:mad}).
If the system being simulated lies on the linewidth-size
relation, then its mass is $M\la M_\lws=25\; M_\odot$. As shown
in \S \ref{sec:scalevlws}, this inequality also holds if the system
has a linewidth greater than that given by the linewidth-size relation.
Similarly, equation (\ref{eq:load}) shows that the size of the system
decreases with $\spcs$; as a result, if the system being simulated
lies on or above the linewidth-size relation, then its size is 
no larger than the size corresponding to the turbulence-dominated
linewidth-size relation, $\ell_0\la 0.4$~pc.
Reference to equations (\ref{eq:load}) and (\ref{eq:mad}) shows that simulations of larger
regions would require higher Mach numbers, given that the \alfven Mach
number is observed to be of order unity.

\section{Conclusions}

Ambipolar diffusion is a key process in molecular clouds
since it redistributes magnetic flux and damps waves.
The importance of ambipolar 
diffusion in a turbulent medium
on a length scale $\ell$ and velocity disperion $v$
is governed by the AD Reynolds number
$\rad=\ell/\lad$, where $\lad=\va^2 t_{ni}/v$ is the length scale over which the magnetic
field must vary in order to have a drift velocity $v$ between the neutrals and ions
\citep{zwe97,zwe02}. 
[Note that $\rad$ is useful in describing 
ambipolar diffusion whenever the velocity field includes a significant turbulent
component; it is not 
useful for non-turbulent, AD-driven gravitational collapse \citep{mou87} or C-shocks
\citep{dra80}, where $\rad$ is of order unity.]
We have carried out two-fluid 
simulations of isothermal, turbulent boxes using the code
ZEUS MPAD (described in LMK) at a resolution of $512^3$ 
for AD Reynolds numbers ranging from $\rad=0.12$ to $\rad=1200$, plus a
simulation with ideal MHD.
The resolution we 
have used is sufficient to resolve the sonic length within the 
inertial range, permitting accurate simulations for our $\calm=3$ calculations.
 The mean Mach numbers were fixed
at $\calm=3$ and $\ma=0.67$, corresponding to a plasma-$\beta$ parameter
$\beta=0.1$. 
The purpose of our simulations was to determine how the properties of the clumps
formed in molecular clouds
depend on $\radlo$. One of the simulations (with $\radlo=12$)
was in the middle of the observed range of the observed
values of the AD Reynolds number; two of the simulations (those with
$\radlo-1.2$ and 120) were somewhat below and above the observed values
of $\radlo$; and the remaining two simulations, with $\radlo=0.12$ and 1200,
were designed to show the transition to hydrodynamics and ideal MHD, respectively.
In order to carry out these simulations, we
used the Heavy Ion Approximation with an ionized mass fraction of $10^{-2}$
(LMK) to represent physical systems with actual ionized mass fractions 
$\sim 10^{-6}$. We validated our simulations with convergence studies at lower
resolution. The power spectra in our simulations
show that the inertial range of our simulations extends down to a length
scale $\ell_0/20$, which is comparable to the sonic length;
it is important to resolve the sonic length in simulations of turbulent boxes since
the density has significant fluctuations on larger scales.
Our principal conclusions are:

\begin{itemize}

\item[1.] Values of the AD Reynolds number $\rad$ in a sample of 15 
molecular clumps with measured magnetic
fields \citep{cru99} range from 3 to 73; the logarithmic mean value is 17.
Omitting one outlier, the clumps with upper limits on the magnetic field
have an average lower limit of 
$\avg{\rad}_{\log} > 18$.
The predicted value of the AD Reynolds number 
for self-gravitating molecular clouds and clumps
with the fiducial ionization is $\rad\simeq 20$, in excellent agreement with
observation.

\item[2.] Several regimes of ambipolar diffusion can be identified, depending on
the ratio of the flow time, $t_f$, to the ion-neutral collision time, $t_{in}$, and the
neutral-ion collision time, $t_{ni}$: (I) ideal MHD 
($t_f/t_{ni}\rightarrow\infty$, corresponding to $\rad\rightarrow \infty$
for a given value of $\ma$); 
(II) standard ambipolar diffusion, with
$t_f>t_{ni}$, so that the neutrals and ions are coupled together over a 
flow time; (III) strong AD ($t_{ni}>t_f>t_{in}$), so that the neutrals are not coupled
to the ions over a flow time, but the ions are coupled to the neutrals; (IV) weakly coupled
($t_{in}>t_f$), so that the ions and neutrals behave almost independently over a flow time;
and (V) hydrodynamics 
($t_f/t_{in}\rightarrow 0$ or $\chi_i\rightarrow 0$, corresponding to
$\rad\rightarrow 0$). The molecular clumps in Crutcher's sample
are all in the second regime, standard AD.

\item[3.] Implied self-gravity:
Since the ionization scales approximately as the square root of the density,
the ambipolar diffusion time is proportional to the gravitational free-fall time \citep{mou87}.
As a result, any simulation of ambipolar diffusion has a gravitational virial parameter
$\avir$ that is determined by $\rad$, $\ma$ and 
the parameters describing the ion-neutral coupling and
the ionization [$\avir\propto (\gad^*\chiis/\rad)^2\ma^4$---eq. \ref{eq:avirtext}]. 
It is not possible to carry out a simulation in which the effects of
self-gravity and ambipolar diffusion are varied independently unless the ionization
is treated as a free parameter.

\item[4.] Clump mass spectrum. 
Using Clumpfind \citep{wil94}, we
found all the clumps with densities exceeding the mean density in the box.
We find that the slope of the higher-mass portion
of the resulting clump mass spectrum increases as $\rad$ decreases, which is qualitatively
consistent with Padoan et al's (2007) finding that the mass spectrum in hydrodynamic
turbulence is significantly steeper than in ideal MHD turbulence. 
The value of the slope that we find for $\rad=12$, the case closest to the value 
observed in molecular clouds, is $\gfit=1.43\pm0.10$, which is consistent with the
Salpeter value, $\Gamma=1.35$. The almost-ideal MHD case
($\rad=1200$) has a slope $\gfit=1.22\pm 0.11$, which is marginally consistent with the Salpeter value.
We further confirm Padoan et al's (2007) relation between the index of the power spectrum and the slope of the clump mass spectrum in the limiting cases of ideal MHD and near hydrodynamics.
However, the value we find for the spectral index in our ideal MHD simulation differs
from theirs, presumably because our simulation has lower values of $\beta$ and $\calm$.
This suggests that the IMF in the \citet{pad02,pad07} turbulent fragmentation model
depends on the environment, which could conflict with evidence for an IMF that is approximately universal (see also \citealp{bal06}).

\item[5.] 
Ambipolar diffusion affects the mass-to-flux ratio of clumps, even
in the absence of self-gravity:
The average mass-to-flux ratio \mfr\ at low $\rad$
is slightly larger than at high $\rad$, and 
the dispersion in the values of \mfr\ for individual clumps at moderate $\rad$ is almost twice that at high $\rad$.

\item[6.] Scaling relations for simulations of isothermal turbulent boxes. A simulation of an isothermal, magnetized, turbulent box is characterized by three dimensional parameters: the size of the box, $\ell_0$, the mean density in the box, $\bar\rho\propto \bar\nh$, and the sound speed, $c_s$ (e.g., \citealp{ost99}).
A single simulation with ideal MHD applies to an infinite range of values
of each of these dimensional parameters, provided that the dimensionless parameters
describing the simulation (in this case, the Mach numbers $\calm$ and $\ma$) are
the same (e.g., \citealp{pad99}). 
Except in regions of high-mass star formation, molecular gas
generally has a temperature $T\sim 10-20$~K, so that $c_s$ is nearly constant; as
a result, there are only two dimensional parameters that have a significant variation,
$\ell_0$ and $\bar\rho$. Each physical process that is introduced into the simulation,
such as ambipolar diffusion, introduces a dimensionless parameter, such as $\rad$,
which must be fixed for the simulation, thereby reducing the number of scaling parameters
by one. For simulations with ambipolar diffusion, the physical parameters
describing the system being simulated are characterized by a single dimensional
parameter (for constant $c_s$), which we took to be the mean density (\S \ref{sec:adscale}).
Such simulations are intrinsically scale free. Even if one includes self-gravity,
then, so long as the ionization scales as $\bar\chi_i\propto \nbh^{-1/2}$, the
simulation remains scale free. However,
if one further requires that the simulation satisfy an observed linewidth-size relation,
then the mean density is determined and there are no independent scaling parameters.

\item[7.] Physical parameters associated with the simulations. 
Two of the simulations we carried out were for the purpose of studying
the transition to ideal MHD [$\radlo=1200$] and to hydrodynamics [$\radlo=0.12$],
so we focus on the remaining three, with $\radlo=1.2,\,12,\,120$.
As discussed
in item (3) above, a simulation
of ambipolar diffusion has an associated value of the virial parameter.
The virial parameter cannot be significantly less than unity since the
self-gravity would induce collapse that would lead to $\avir\sim 1$.
This constraint gives a lower bound on the ionization such
that $\radlo\la 6.4\gad^*\chiis$. As a result, two of the simulations
[$\radlo=12,\;120$] could be realized in nature only if the
ionization and/or the ion-neutral coupling constant
were larger than the fiducial values, which correspond
to $\gad^*\chiis=1$. The $\radlo=1.2$ simulation corresponds to a system
in which self-gravity is unimportant (for $\gad^*\chiis=1$),
whereas the $\radlo=12,\,120$ simulations correspond to
systems that mostly likely are gravitationally bound 
(for $\gad^*\chiis$ as close to unity as possible).
Because of this constraint and because of the small
value of the Mach number we adopted ($\calm=3$), our simulations apply to
small regions in molecular clouds, with $\ell_0\la 1/\nbht^{1/2}$~pc
and $M_0\la 100/\nbht^{1/2}\;M_\odot$. 
If the system being simulated 
has a velocity dispersion on or above the linewidth-size relation observed
in the Galaxy, then the size
of the region is $\ell_0\la 0.4$~pc and the mass is $\la 25 M_\odot$. 

\item[8.] A general discussion of
scaling relations for self-gravitating systems is given in the Appendix.
In applying the  linewidth-size relation, we follow
\citet{fal10} in distinguishing the turbulence-dominated relation
from the virialized one.

\end{itemize}

\acknowledgments

We thank Charles Hansen, Patrick Hennebelle,
Mark Heyer, Mark Krumholz, Enrique Vazquez-Semadeni,
Ellen Zweibel, and particularly an anonymous referee
and Telemachos Mouschovias for helpful comments.
This research has been supported by the NSF under grants AST-0606831 
and AST-0908553  and by NASA under an ATFP grant,
NNX09AK31G. CFM also acknowledges the support of the Groupement d'Int\'er\^et Scientifique
(GIS) ``Physique des deux infinis (P2I)" at the completion of this work.
RIK received support for this work provided by the 
US Department of Energy at Lawrence Livermore National Laboratory 
under contract DE-AC52-07NA 27344.
This research was also supported by the grant of high performance computing resources from the National Center of Supercomputing Application through grant TG-MCA00N020.

\appendix

\section{Scaling Laws for Isothermal Turbulent Boxes}

In this Appendix, we give a general discussion of scaling laws for simulations
of isothermal, turbulent gases in a box. Although we do not include the
effects of self-gravity in the text, we do include it here, so as to make
the discussion more generally useful. We focus on molecular gas, since such
gas is generally approximately isothermal. 
Particular scaling relations that have been derived previously are noted 
\citep{kle00,ost99,til04,vs05a,vs08}.

In the simplest case in which there is no gravity and the MHD is ideal,
a simulation of an isothermal, magnetized, turbulent box is characterized by
two dimensionless parameters \citep{pad99}, the 3D sonic Mach number, 
$\calm=3^{1/2}\snt/c_s$,
and the plasma-$\beta$ parameter,
$\beta\equiv 8\pi\bar\rho c_s^2/\brms^2$. Here $\snt$ is the 1D nonthermal 
velocity dispersion, $c_s$ the isothermal sound speed, $\bar\rho$ the
mean mass density and 
$\brms\equiv\avg{B^2}^{1/2}$ the rms magnetic field. 
Equivalently, the two parameters can be chosen to be the
sonic Mach number and the Alfven Mach number, $\ma\equiv 3^{1/2}\snt/v_{\rm A}$,
since the plasma-$\beta$ parameter is related to $\calm$ and $\ma$ by
\beq
\beta=2\left(\frac{\ma}{\calm}\right)^2.
\label{eq:beta}
\eeq
In general, the Mach numbers, $\calm$ and $\ma$, and the plasma-$\beta$ 
parameter are functions of time.

The turbulent box is also characterized by
three dimensional parameters: the size of the box, $\ell_0$, the mean density
in the box, $\bar\rho=M_0/\ell_0^3$, where $M_0$
is the mass in the box, and the isothermal sound speed, $c_s$
(e.g., \citealp{ost99}).
In the absence of
other physical processes, these parameters can be selected arbitrarily. 
In other words, a given simulation corresponds to definite values  of
$\calm$ and $\beta$, but it can be scaled to arbitrary values of
$\bar\rho$, $\ell_0$, and $c_s$. However, the introduction of a 
new physical process, such as self-gravity or ambipolar diffusion, 
introduces a new
dimensional constant and a corresponding new dimensionless parameter, so
that the number of independent dimensional parameters is reduced by one.
The same reduction occurs if a relation between dimensional parameters is
assumed, such as a relation between the size of the box and the mean velocity
dispersion (a linewidth-size relation). 
In many cases, the temperature is tightly constrained, so that in fact
there are only two dimensional parameters that can be chosen at will. Hence, if an 
isothermal system satisfies a linewidth-size relation and is either
self-gravitating or subject to ambipolar diffusion, then
its velocity scale $c_s\propto T^{1/2}$ is set
by the assumed temperature and its size and mean density are
determined by dimensionless parameters; in this case a given
simulation applies to only a single set of parameters describing the box.

	We return to the simplest case in which there is neither self-gravity nor 
ambipolar diffusion. Interstellar densities are often given in terms of number
densities; we use the
the density of hydrogen nuclei, $\nh=\rho/\muh$, where $\muh$ is the mass per hydrogen
nucleus ($=2.34\times 10^{-24}$~g for cosmic abundances).
Numerically, we have
for the mass, flow time, and column density of the box,
\begin{eqnarray}
 M_0&=&\bar\rho \ell_0^3~=~34.6\nbht\lpc^3~~~M_\odot,
 \label{eq:mo}\\
t_f&\equiv&\frac{\ell_0}{v_{\rm rms}}=\frac{\ell_0}{\calm c_s}= 5.19\times 10^6\left(\frac{\lpc}{\calm T_1^{1/2}}\right)~~~\mbox{yr},\\
\Nh&=&\nbh\ell_0=3.09\times 10^{21}\nbht\lpc~~~\mbox{cm}^{-2},
\end{eqnarray}
where $T_1\equiv T/(10$~K) 
and $c_s=0.188\,T_1^{1/2}$~km~s\e\ for molecular gas with cosmic abundances.
The column density corresponds to a surface density 
\beq
\Sigma=2.34\times 10^{-3} N_{\rm H,\, 21}~\mbox{g cm\ee}=11.2
N_{\rm H,\, 21}~M_\odot~\mbox{pc\ee},
\eeq
where $N_{\rm H,\, 21}\equiv \Nh/(10^{21}$~H~cm\ee).
The visual extinction corresponding to this column is 
$A_V=N_{\rm H,\, 21}\delta$~mag,
where $\delta$ is the ratio of the extinction per unit mass to the Galactic value.
The magnetic field is given by 
(\citealp{ost99}; note that their $\beta$ is half the normal value)
\beqa
\brms&=&\left(\frac{8\pi\bar\rho c_s^2}{\beta}\right)^{1/2}=
4.56\;\left(\frac{\nbht T_1}{\beta}\right)^{1/2}~~\mug,\\
	&=&3.23\;(\nbht T_1)^{1/2}\frac{\calm}{\ma}~~\mug.
\label{eq:brms}
\eeqa

\subsection{Scaling Relations for MHD Simulations of Turbulent Boxes with Self Gravity}

As discussed above, self-gravity introduces an additional dimensionless parameter
into a simulation and therefore reduces the number of independent dimensional parameters
by one. For the case in which the ionization scales as $\bar\chi_i\propto \nbh^{-1/2}$,
this reduction is the same as that due to the inclusion of ambipolar diffusion
(\S \ref{sec:adsg});
that is, for such an ionization law, simulations with both self gravity and
ambipolar diffusion obey the same scaling relations as simulations with only
one of these processes. 
In this section, we first summarize the dimensionless parameters
used to characterize turbulent simulations with self gravity. We then describe variants of the
Jeans mass that take into account turbulent motions. Finally, scaling laws for
self-gravitating systems are given for turbulent boxes ($c_A=c_V=1$---see
eqs \ref{eq:ca} and \ref{eq:cv}). With two dimensional parameters specified---the
strength of self gravity and the temperature---there is still one free dimensional parameter;
as a result, self-gravitating, magnetized turbulent boxes are scale free. 

\subsubsection{Dimensionless parameters}
\label{sec:dimless}

There are several equivalent 
dimensionless parameters that can describe the effects of self-gravity. One is
the ratio of the mass to the characteristic mass of a self-gravitating cloud,
$c_s^3/(G^3\bar\rho)^{1/2}$:
\beq
\mu_0\equiv \frac{ M_0}{c_s^3/(G^{3/2}\bar\rho^{1/2})}.
\eeq
This parameter is related to the mass, length and sound speed by
\beq
\frac{G M_0/\ell_0}{c_s^2}=\mu_0^{2/3}.
\label{eq:twothirds}
\eeq
In terms of the free-fall velocity, $\vffo\equiv(G M_0/2R)^{1/2}$, we have
$\mu_0=(\vffo/c_s)^3$. 

Another parameter describing the effects of self-gravity is the number of Jeans lengths in the
box \citep[e.g.][]{ost99}. The typical Jeans length 
in the box is $\lj=(\pi c_s^2/G\bar\rho)^{1/2}$, so
the number of Jeans lengths in the box is 
\beq
\nj\equiv \frac{\ell_0}{\lj}=\frac{\mu_0^{1/3}}{\surd\pi}.
\eeq
The Jeans mass for the box is 
usually defined as
$M_{\rm J}\equiv\bar\rho\lj^3$, so that the number of Jeans masses in the box is
$M_0/\mj=(\ell_0/\lj)^3=\nj^3$.
The corresponding value
of $\mu$ is $\mu_{\rm J}=\mj(G^3\bar\rho)^{1/2}/c_s^3=
\pi^{3/2}\simeq 5.57$.

A third parameter describing the effects of self-gravity is the virial parameter,
which for a spherical cloud of radius $R$ is
\beq
\avir=\frac{5\sigma^2R}{G M_0}
\label{eq:avirr}
\eeq
\citep{ber92}; here $\sigma^2=c_s^2+\snt^2$ is the total
1D velocity dispersion. Self gravity is important for $\avir\simeq 1$ and is
unimportant for $\avir\gg 1$. By contrast, $\mu_0$ and $n_J$ can
have arbitrary values $\ga 1$ when self gravity is important. 
In further contrast to $\mu_0$ and $n_J$, the effects of
bulk kinetic energy as well as thermal energy are included in $\avir$.
For gas in a box, we define $\avir$ by replacing
$R$ by $\ell_0/2$:
\beq
\avir\equiv \frac{5\sigma^2 \ell_0}{2G M_0},
\label{eq:avir}
\eeq
which is the same as equation (\ref{eq:avir1}) in the text.
There is a complication here, since $\avir$ is defined with respect to the total
velocity dispersion, $\sigma$, whereas the linewidth-size relation depends only on
the non-thermal velocity dispersion, $\snt$.
For $\calm^2\gg 1$, there is no problem, since the two velocity dispersions are nearly
the same. Relations involving the Mach number that do not depend on
the linewidth-size relation can be extended to low Mach numbers by redefining
$\calm$ as $[3(1+\snt^2/c_s^2)]^{1/2}$; otherwise, such relations are restricted to
$\calm^2\gg 1$. Bearing this in mind, we note that
equation (\ref{eq:twothirds}) implies that $\avir$ is related to the other two parameters by
\beq
\avir=\frac 56\left(\frac{\calm^2}{\mu_0^{2/3}}\right) = \frac{5}{6\pi}\left(\frac{\calm^2}{n_J^2}\right).
\label{eq:avirmu}
\eeq
\citet{vs08} derived a similar expression for spherical clouds (their
$\alpha=\frac 35 \avir$ and their $\nj$ is $R/\lj$, which is half the value we use).
The virial parameter is also related to the ratio
of the Jeans length to the sonic length (\S \ref{sec:sonic}):
Equations (\ref{eq:machsonic}) and (\ref{eq:avir}) give the relation between the
virial parameter and the sonic length as
\beq
\avir=\frac{5}{6}\frac{c_s^2}{G\bar\rho\calm^2\ell_s^2}
\eeq
for $q=\frac 12$.
It follows that the ratio of the Jeans length to the sonic length is
\beq
\frac{\lj}{\ell_s}=\left(\frac{6\pi\avir}{5}\right)^{1/2}\calm.
\eeq

The parameters that describe the effects of self-gravity determine the
ratio of the flow time to the free-fall time, which is
\beq
t_{\rm ff}=\left(\frac{3\pi}{32G\bar\rho}\right)^{1/2}=1.37\times 10^6\nbht^{-1/2}~~~\mbox{yr}.
\label{eq:tff}
\eeq
Relative to the free-fall time, the flow time is 
\beq
\frac{t_f}{t_{\rm ff}}=1.84\;\frac{\mu_0^{1/3}}{\calm}=3.27\;\frac{\nj}{\calm}=
	\frac{1.68}{\avir^{1/2}}.
\label{eq:tftff}
\eeq

\subsubsection{Variants of the Jeans length and Jeans mass}

We can define both large-scale and small-scale variants of the Jeans length and
Jeans mass. On large scales, the density is close to the mean, $\bar\rho$, but
the velocity dispersion is
$\sigma\equiv(\snt^2+c_s^2)^{1/2}=\calm c_s/\surd 3$. We therefore define
the ``turbulent" variants of the Jeans length and Jeans mass 
by replacing the sound speed $c_s$ with the velocity dispersion $\sigma$,
\beqa
\ljt &\equiv& \left(\frac{\pi \sigma^2}{G\bar\rho}\right)^{1/2}=\left(\frac{\calm}{\surd 3}\right)\lj,\\
\mjt&\equiv& \bar\rho\ljt^3=\left(\frac{\calm}{\surd 3}\right)^3 \mj.
\eeqa
The corresponding dimensionless quantities are
\beqa
\mu_{\turb,\,0} &\equiv& \frac{ M_0}{\sigma^3/(G^{3/2}\bar\rho^{1/2})}=\left(\frac{\surd 3}{\calm}
	\right)^3 \mu_0,\\
\njt&\equiv&\frac{\ell_0}{\ljt}=\left(\frac{\surd 3}{\calm}\right) \nj.
\eeqa
Both $\mu_{\turb,\,0}$ and $\njt$ are of order unity when the virial parameter is:
\beq
\avir=\frac{5}{2\mu_{\turb,\,0}^{2/3}}=\frac{5}{2\pi\njt^2}.
\eeq

	On small scales, however, the velocity dispersion is about equal to the
sound speed, $c_s$, whereas the density can vary over orders of magnitude.
In star-forming cores, the typical pressure is 
the mean turbulent pressure
$\bar\rho\snt^2$ \citep{pad95,kru05};
for an isothermal gas, this corresponds to a density 
$\rhocore\equiv (\calm^2/3)\bar\rho$.
We now introduce another variant of the Jeans length,
the ``core Jeans length," $\ljc$, in which the velocity dispersion and density
are those expected in star-forming cores,
\beq
\ljc=\left(\frac{\pi c_s^2}{G\rhocore}\right)^{1/2}=\left(\frac{\pi c_s^2}{G\calm^2\bar\rho/3}\right)^{1/2}
=\left(\frac{\surd 3}{\calm}\right)\lj.
\label{eq:ljc}
\eeq
For supersonic flows, the core Jeans length is indeed small compared
to the turbulent Jeans length, 
$\ljc=(3/\calm^2)\ljt$, since it measures the effect of high pressures on thermally supported
gas, whereas $\ljt$ measures the effect of the turbulence on gas at the
average density. 
If self-gravity is important in the turbulent box ($\avir\sim 1$),
the core Jeans length is somewhat greater than the sonic length,
\beq
\frac{\ljc}{\ell_s}=\left(\frac{18\pi\avir}{5}\right)^{1/2}=3.36\avir^{1/2}.
\eeq
The ``core Jeans mass" is smaller than the normal Jeans mass and much
smaller than the turbulent Jeans mass,
\beq
\mjc=\rhocore\ljc^3=\frac 13 \calm^2\bar\rho \ljc^3 = \left(\frac{\surd 3}{\calm}\right) \mj.
\label{eq:mjcore}
\eeq
We expect $\mjc$ (or perhaps the somewhat smaller core Bonnor-Ebert mass)
to be the typical mass of gravitationally bound cores in a turbulent cloud \citep{pad95}.

\subsubsection{Scaling in Terms of the Mean Density}
\label{sec:gscale}

As discussed at the outset, the introduction of an additional physical process, such as
self-gravity, reduces the number of independent dimensional parameters to two, which
we take to be the density $\nh$ and the temperature $T$. Since the temperature
has little variation in molecular clouds, there is effectively only one independent
dimensional parameter, $\nh$. In terms of $\nh$, $T$ and 
the dimensionless parameters describing the
self gravity, the size and mass of the box are then given by
\begin{eqnarray}
\lpc&=&0.488\mu_0^{1/3}\left(\frac{T_1}{\nbht}\right)^{1/2}
=0.865\nj\left(\frac{T_1}{\nbht}\right)^{1/2}
=0.445\;\frac{\calm}{\avir^{1/2}}\left(\frac{T_1}{\nbht}\right)^{1/2},
\label{eq:lgrav}\\
\frac{ M_0}{M_\odot}&=&4.01\left(\frac{\mu_0 T_1^{3/2}}{\nbht^{1/2}}\right)
=22.3\left(\frac{\nj^3 T_1^{3/2}}{\nbht^{1/2}}\right)
=3.05\left(\frac{\calm^3 T_1^{3/2}}{\avir^{3/2}\nbht^{1/2}}\right),
\label{eq:mgrav}\\
N_{\rm H,\, 21}&=&1.50\mu_0^{1/3}(\nbht T_1)^{1/2}
=2.67\nj(\nbht T_1)^{1/2}
=1.37\left(\frac{\nbht T_1}{\avir}\right)^{1/2} \calm.
\label{eq:Ngrav}
\end{eqnarray}
Note that the column density is directly proportional to the square root of
the thermal pressure in the first two cases, and to the square root of the 
turbulent pressure, $\bar\rho v_\rms^2\propto \calm^2 \nbh T$,
 in the third case; this is to be expected, since the pressure in
a self-gravitating system is proportional to $G\Sigma^2$.
The flow time is given by equations (\ref{eq:tff}) and (\ref{eq:tftff}),
the magnetic field by equation (\ref{eq:brms}) and the ratio of the mass
to the magnetic critical mass, $\mu_{\Phi,\,0}$, by equation (\ref{eq:muphi}).
The scaling relation between the size and density in terms of $\nj$ has been given by
\citet{ost99} and \citet{vs05a}; for the mass in terms of the size and $\nj$ by
\citet{til04}, although they have a different numerical coefficient
than implied by the above relations; and by \citet{kle00}
for both the mass and the size in terms of the density for the particular case they
consider, which has $\nj=4$. 

The Jeans mass and the Bonnor-Ebert mass are
\beq
\mj=22.3\left(\frac{T_1^{3/2}}{\nbht^{1/2}}\right)~~~M_\odot,~~~~~~
M_{\rm BE}=4.74\left(\frac{T_1^{3/2}}{\nbht^{1/2}}\right)~~~M_\odot,
\label{eq:mjbe}
\eeq
where for the Jeans mass we have assumed that
$\nbh$ is the mean density in the ambient medium, and for the Bonnor-Ebert mass
we have assumed that $\nbh$ is 
the density at the surface of the Bonnor-Ebert sphere.
Note that $\mj$ and $\mbe$ can be expressed in terms of the surface density
of the box as, for example, by
\beq
\mj=3.36\left(\frac{\mu_0^{1/3}T_1^2}{N_{\rm H\, 22}}\right)~~~M_\odot,~~~~~~
M_{\rm BE}=0.713\left(\frac{\mu_0^{1/3}T_1^2}{N_{\rm H,\, 22}}\right)~~~M_\odot.
\eeq
We have changed the normalization of the column density so as to yield values
of the Jeans mass and Bonnor-Ebert mass comparable to observed values for
$\Nhtt\equiv\Nh/(10^{22}\mbox{ cm\ee})\sim 1$.
The core values are smaller by a factor $\surd 3/\calm$. In this case, it is
convenient to express the results in terms of the virial parameter of the box,
\beq
M_{\rm J,\,core}=5.31\left(\frac{T_1^2}{\avir^{1/2} N_{\rm H,\, 22}}\right)~~~M_\odot,~~~~~~
M_{\rm BE,\, core}=1.13\left(\frac{T_1^2}{\avir^{1/2} N_{\rm H,\, 22}}\right)~~~M_\odot.
\eeq
Note that in both cases, the critical masses have no explicit dependence on the Mach number.

The basic conclusion is that MHD simulations of
of self-gravitating, turbulent boxes are scale free; even with the temperature
fixed, there is one free parameter, such as the box size or the density, that
can be chosen arbitrarily. Similarly, MHD simulations with ambipolar diffusion
are scale free, as we have seen in \S \ref{sec:scaling}. If the ionization
scales as $1/\nbh^{1/2}$, then $\rad$ and $\avir$ are directly related
(\S \ref{sec:adsg}) and simulations with both ambipolar diffusion and
self gravity are also scale free.

\subsection{Scaling with the Linewidth-Size Relation}

\subsubsection{Linewidth-Size Relations for Molecular Clouds and Turbulent Boxes}
\label{sec:lwsrelations}

Molecular gas in the Galaxy exhibits a linewidth-size relation in which
the velocity dispersion of the gas increases as a power of the physical dimension of the
region (Larson 1981). Heyer \& Brunt (2004) have shown that this applies
within GMCs as well as among different GMCs. The data are consistent
with the relation
for the 1D velocity dispersion
\beq
\snt=\spc R_{\rm pc}^q,
\label{eq:lws}
\eeq
with $\spc=0.72$~km s\e\ and $q=\frac 12$ \citep{sol87,mck07};
for these parameters, the Mach number is
$\calm=6.63(R_{\rm pc}/T_1)^{1/2}.$
Such a relation appears to be satisfied by most molecular gas in
the Galaxy; for example, Heyer \& Brunt (2004) find $q=0.49\pm0.15$ within
individual molecular clouds in their sample, and
\citet{fal09} find that the combined data from several different surveys shows
a clear linewidth-size relation, although they do not give a fit.
The parameters have different values in regions of high-mass star formation, however:
in such regions, $\spc\sim$ a few km s\e,
different regions of high-mass star formation
do not have line widths that increase as $R^{1/2}$, and it is
not known how the velocity dispersion within individual regions scales with size
(Plume et al 1997). 
Nonetheless, equation (\ref{eq:lws}) 
appears to be satisfied in these
regions to within an order of magnitude.

\citet{mou95} and \citet{hey09} have proposed variants of the linewidth-size relation
for gas that is gravitationally bound. 
\citet{fal10} have reconciled these {\it virialized linewidth-size relations}
with the classical {\it turbulence-dominated linewidth-size relation}.
Recall that the surface density of the 
cloud is $\Sigma_0=M_0/(c_A\ell_0^2)$, where $c_A=(1,\pi/4)$ for a box and
a spherical cloud, respectively. The virial parameter (eq. \ref{eq:avir}) is then
\beq
\avir\equiv\frac{5\sigma^2}{2Gc_A\Sigma_0\ell_0},
\label{eq:avirsigma}
\eeq
where we have used the identity symbol to emphasize that this follows directly
from the definitions of the quantities involved; there is no physics in this relation.
Solving this relation for the velocity dispersion gives
\beq
\sigma\equiv\left[\frac{\pi}{5}\left(\frac{c_A}{\pi/4}\right) G\avir \Sigma_0 R_0\right]^{1/2},
\label{eq:sigmavir}
\eeq
where $R_0\equiv \ell_0/2$ is the cloud radius. The factor in parentheses with
$c_A$ is unity for a spherical cloud. This relation is superficially  like the linewidth-size
relation in equation (\ref{eq:lws}), but it is quite different: First, the exponent in
the linewidth-size relation, $q\simeq \frac 12$, follows from observation, whereas that in
equation (\ref{eq:sigmavir}) is $\frac 12$ by definition, and second the
coefficient in relation (\ref{eq:sigmavir}) depends on the column density.
\citet{hey09} inserted the physics into this relation by noting that
gravitationally bound clouds have $\avir\simeq 1$. 
For spherical clouds, they then found
\beqa
\sigma&=&\left[\frac{\pi}{5}G \Sigma_0 R_0\right]^{1/2},
\label{eq:sigmavir1}\\
&=& 0.55(\Nhtt R_\pc)^{1/2}~~~\mbox{km s\e}
\label{eq:sigmavir1n}
\eeqa
which we term the virialized linewidth-size relation.
They obtained a sample
of bound clouds by combining the $^{12}$CO data on the \citet{sol87} molecular
clouds with data from the higher resolution $^{13}$CO data on these
clouds from the Galactic Ring Survey (REF). Over a range of surface
densities $10\,M_\odot~\mbox{pc\ee}\la\Sigma\la 10^3\,M_\odot$~pc\ee,
equation (\ref{eq:sigmavir1n}) describes the data well, after a somewhat uncertain
correction is made for the cloud masses. \citet{mou95} previously found an analogous
relation for magnetized clouds with $B\mu_{\Phi,\,0}$ in place of $\Nh$ (see
eq. \ref{eq:muphi}).

\citet{fal10} concluded that non-self gravitating interstellar
gas obeys the turbulence-dominated linewidth-size relation given by
equation (\ref{eq:lws}), whereas self-gravitating gas satisfies the virialized
linewidth-size relation given by equation
(\ref{eq:sigmavir1}). There is 
a critical surface density that defines the boundary between the turbulent
and virialized cases: Equating the velocity dispersions for the two cases gives
\beqa
\Sigma_\lws&=&\frac{5}{\pi G}\left(\frac{\pi/4}{c_A}\right)\left(\frac{\spc^2}{1\;\mbox{pc}}\right),\\
&=& 192\left(\frac{\pi/4}{c_A}\right)\spcs^2~~~M_\odot\;\pc^2,
\eeqa
where $\spcs=\spc/(0.72$~km~s\e) is normalized to the standard Galactic
value (see eq. \ref{eq:spcs}) 
and $c_A=\pi/4$ for spherical clouds. The corresponding column density 
is
\beq
N_\lws=1.71\times 10^{22}\left(\frac{\pi/4}{c_A}\right)\spcs^2~~~\mbox{cm\ee}. 
\label{eq:nlws}
\eeq
It should be noted that
these values for $\Sigma_\lws$ and $N_\lws$ for spherical clouds
are comparable to the mean values for Galactic
GMCs found by \citet{sol87}, and about twice the mean values found
by \citet{hey09}. Regions with $\Sigma\ll\Sigma_\lws$ are
dominated by interstellar turbulence, whereas those with
$\Sigma\gg\Sigma_\lws$ are dominated by self-gravity and
are decoupled from the turbulent cascade in the interstellar medium.
The high surface densities of regions of high-mass star formation thus
naturally lead to the high velocity dispersions observed there by,
for example, \citet{plu97}.
The maximum density for a cloud satisfying the turbulent linewidth-size relation
is
\beq
\bar n_\lws=\left(\frac{c_A}{c_V}\right)\frac{N_\lws}{\ell}=1.83\times 10^5\left(\frac{\pi/6}{
c_V}\right)\frac{\spcs^4}{\calm^2 T_1}~~~\mbox{cm\eee}.
\label{eq:nblws}
\eeq
where we used equation (\ref{eq:calm}) to eliminate $\ell$ and the relations
$M\propto c_A \Nh \ell^2=c_V \bar\nh\ell^3$ to cover the different geometries.
The maximum mass for a cloud satisfying the turbulent linewidth-size relation
is
\beq
M_\lws=0.311\left(\frac{\calm^4 T_1^2}{\spcs^2}\right)~~~M_\odot,
\label{eq:mlws}
\eeq
which is independent of geometry.

\subsubsection{Turbulent Boxes with the Turbulence-Dominated Linewidth-Size Relation
{\rm (}$\Nh\leq N_\lws$, $c_A=c_V=1${\rm )}}
\label{sec:scalelws}

Here we determine the scaling relations for turbulent boxes 
that satisfy the turbulence-dominated linewidth-size relation
\beq
\snt=0.72\spcs R_\pc^{1/2}~~~\mbox{km s\e}.
\eeq
The Mach number $\calm$ and box size $\ell_0$ are related by
equation (\ref{eq:calm}). As a result,
the properties of the turbulent box are given by:
\begin{eqnarray}
\ell_0&=& \frac 23\;\frac{\calm^2 c_s^2}{(\spc^2/\mbox{ 1pc})}=0.0455\left(\frac{\calm^2 T_1}{\spcs^2}\right)~~\mbox{pc},
\label{eq:lwslo}\\
t_f&=&2.36\times 10^5\left(\frac{\calm T_1^{1/2}}{\spcs^2}\right)~~~\mbox{yr},
\label{eq:lwstf}\\
 M_0&=& 3.25\times 10^{-3}\left(\frac{\calm^6\nbht T_1^3}{\spcs^6}\right)~~M_\odot,\\
\Nh&=&1.40\times 10^{20}\left(\frac{\calm^2\nbht T_1}{\spcs^2}\right)~~~\mbox{cm\ee}.
\label{eq:Nhlws}
\end{eqnarray}

For simulations that include self-gravity but have surface densities less than
the critical one ($\Sigma\leq\Sigma_\lws$), 
the column density is most simply expressed in terms of the 
linewidth-size parameter and the virial parameter
using equations (\ref{eq:avirsigma}) and (\ref{eq:lwslo}), 
\beq
\Nh=\frac 54 \left(\frac{\spc^2}{\mbox{1 pc}}\right)\frac{1}{G\muh\avir}
=1.34\times 10^{22}\left(\frac{\spcs^2}{\avir}\right)~~~\mbox{cm\ee},
\label{eq:Ntlws}
\eeq
which corresponds to $\Nh=N_\lws/\avir$ for $c_A=1$ in equation (\ref{eq:nlws}).
The scaling for the density can be expressed
in terms of the linewidth-size parameter $\spcs$ and a parameter
describing the self gravity with the aid of
equations (\ref{eq:lwslo}) and (\ref{eq:lgrav}) ,
\beq
\nbht=115\left(\frac{\mu_0^{2/3}\spcs^4}{\calm^4 T_1}\right)
=361\left(\frac{\nj^2\spcs^4}{\calm^4 T_1}\right)
=96\left(\frac{\spcs^4}{\avir\calm^2T_1}\right).
\eeq
Comparison with equation (\ref{eq:nblws}) for a box geometry ($c_V=1$) shows
that $\nbh=n_\lws/\avir$, where $\avir\ga 1$. Similarly, equation (\ref{eq:mgrav}) implies
\beq
\frac{ M_0}{M_\odot}=0.374\left(\frac{\mu_0^{2/3}\calm^2T_1^2}{\spcs^2}\right)
=1.17\left(\frac{\nj^2\calm^2T_1^2}{\spcs^2}\right)
=0.311\left(\frac{\calm^4T_1^2}{\avir\spcs^2}\right),
\label{eq:lwsmgrav}
\eeq
so that $M_0=M_\lws/\avir$, where $M_\lws$ is given in equation (\ref{eq:mlws}).
The turbulence-dominated linewidth-size relation does not
apply for densities exceeding $n_\lws$,
corresponding to column densities $\Nh>N_\lws$ and masses
$M>M_\lws$, and for that case the scaling is given by
the relations in \S \ref{sec:scalevlws} below.

When expressed in terms of the linewidth-size relation,
the core values for the Jeans mass and Bonnor-Ebert mass (eq. \ref{eq:mjcore})
are independent of the Mach number,
\beq
M_{\rm J,\, core}=3.96\;\frac{\avir^{1/2}T_1^2}{\spcs^2}~~~M_\odot,~~~~~~
M_{\rm BE,\, core}=0.84\;\frac{\avir^{1/2}T_1^2}{\spcs^2}~~~M_\odot.
\label{eq:mbecore}
\eeq
The core Bonnor-Ebert mass is comparable to the typical mass of observed
stars, particularly if allowance is made for the fact that only a fraction of the
core mass is incorporated into the final star \citep[e.g.][]{mat00}.
These relations can be expressed in terms of the sonic
length instead of $\spcs$ by using equation (\ref{eq:lspc}).

In applying these relations to simulations with driven turbulence, it must be kept in mind that
the driving generally results in deviations from the linewidth-size 
relation (\ref{eq:lws}) on the driving scale.
Using the linewidth-size relation to relate simulations to actual systems is therefore
best done for cases in which the driving is restricted to large scales; the
simulations discussed in the text satisfy this constraint since
they are driven over a narrow range of wavenumbers at the largest scale, 
$1\leq k\leq k_d$, with the driving wavenumber $k_d=2$.
(Here $k$ is a dimensionless wavenumber that is
related to the physical wavenumber $k_{\rm phys}$ by $k\equiv k_{\rm phys}\ell_0/2\pi$;
the minimum possible wavenumber is $k=1$ and the maximum is $\caln_g/2$, where $\caln_g$ is the number of grid cells in each side of the box.) When the turbulence is driven, one must distinguish between the linewidth-size relation
applied to the entire box, and the linewidth-size relation inside the box.
In our simulations, the mean Mach number is approximately constant over the
range $1\leq k\leq k_d=2$. We have chosen to use the full size of the box in
relating our simulations to clouds: $\ell_0=2 R$, where $R$ is the radius of the cloud
or of a region inside the cloud. However, in determining properties inside the
cloud, such as the sonic length, it is necessary to allow for the fact
that the internal linewidth-size relation is normalized approximately to the 
driving scale. As a result, for $\sigma\propto \ell^{1/2}$, the sonic length
is $\ell_s\simeq (\ell_0/k_d)/\calm^2$ rather than $\ell_0/\calm^2$ (see eq. \ref{eq:soniclength}).

\subsubsection{General Scaling with the Linewidth-Size Relation}
\label{sec:scalevlws}

The virialized linewidth-size relation follows from assuming that the
virial parameter is unity, so
the scaling relations for this case are given by the results
in \S \ref{sec:gscale} with $\avir=1$.
To cover both the turbulence-dominated and virialized cases, note
that equation (\ref{eq:Nhlws}) shows that $\Nh\propto\nbh$
for the turbulence-dominated case ($\avir\ga 1$),
corresponding to $\nbh\la \bar n_\lws$, whereas
equation (\ref{eq:Ngrav}) shows that $\Nh\propto \nbh^{1/2}$ for $\avir=1$,
corresponding to $\nbh\ga \bar n_\lws$.
As a result, we have
\beq
\Nh=N_\lws\min\left[\frac{\nbh}{n_\lws},\;\left(\frac{\nbh}{n_\lws}\right)^{1/2}\right]
\label{eq:nsigmalws}
\eeq
for the turbulence-dominated and virialized cases, respectively,
as can be verified by direct substitution using equations (\ref{eq:Ngrav}),
(\ref{eq:nlws}) and (\ref{eq:nblws}). 
Similarly one can show that
\beq
M_0=M_\lws\min\left[\frac{\nbh}{\bar n_\lws},\;\left(\frac{\bar n_\lws}{\nbh}\right)^{1/2}\right]
\label{eq:momlws}
\eeq
with the aid of equations (\ref{eq:mgrav}), (\ref{eq:nblws}) and (\ref{eq:mlws}).
Note that $M_\lws$ is the maximum possible mass for a cloud with a given
velocity dispersion, $\sigma\propto \calm T^{1/2}$, and linewidth-size coefficient,
$\spcs$ \citep{fal10}. Furthermore, for a simulation with a given Mach number,
$M_\lws$ decreases as $\spcs$ increases (eq. \ref{eq:mlws}); as a result, $M_\lws$ is also
the maximum mass of a cloud with a linewidth above the linewidth-size relation.
The size of the simulation box is
\beq
\ell_0=0.0455\left(\frac{\calm^2 T_1}{\spcs^2}\right)
\min\left[1,\;\left(\frac{n_\lws}{\nbh}\right)^{1/2}\right]
~~~\mbox{pc}
\eeq
based on equation (\ref{eq:lgrav}).
It must be borne in mind that these equations are based on the mean linewidth-size
relation; for a given size and/or surface density, the velocity dispersion can vary
by a factor of a few.
Thus the virialized linewidth-size relation, which applies to regions with high column densities
by definition, correspondingly applies to regions of  high density but 
with sizes and masses that decrease as the column density increases.

\subsection{Code Units}
\label{sec:codeunits}

	Numerical codes are generally written in dimensionless form, with
masses, lengths and times written in terms of code units,
$\mtil= M/\mcode$, $\ltil=\ell/\lcode$, and $\ttil=t/\tcode$. 
The code units can be adjusted to fit the problem being simulated.
The properties of the box in code units, $\mtil_0=M_0/\mcode$
and $\ltil_0=\ell_0/\lcode$, can be selected arbitrarily
prior to the simulation (e.g., $\mtil_0=8$ and $\ltil_0=2$), as can 
the normalized sound speed, $\cstil=c_s\tcode/\lcode$. If there are $\caln_g$ grid cells
in each side of the simulation box, the grid size is $\Delta\ltil=\ltil_0/\caln_g$.
For stationary gas, the time step is $\Delta\ttil = C\Delta\ltil/\cstil$, where $C$ is the Courant number.

	The code unit for length is given by $\lcode=\ell_0/\ltil_0$, where
$\ell_0$ is given by equation (\ref{eq:lwslo}) if the
typical Galactic linewidth-size relation is adopted, by
equation (\ref{eq:lgrav}) for a self-gravitating gas, and
by equation (\ref{eq:load}) for a gas undergoing ambipolar diffusion.
The corresponding code unit for time is given by $\tcode=\cstil\lcode/c_s$.
The code unit for mass is given by $\mcode=M_0/\mtil_0$, where
$M_0$ is given by equation (\ref{eq:mgrav}) for a self-gravitating
gas, by equation (\ref{eq:lwsmgrav}) for a self-gravitating gas that
obeys the linewidth-size relation, and by equation (\ref{eq:mad})
for a gas undergoing ambipolar diffusion. 

	The gravitational constant in the code is
\beq
\tilde G=\frac{G\mcode\tcode^2}{\lcode^3}= \left(\frac{G M_0}{\ell_0 c_s^2}\right)\frac
{\tilde\ell_0\cstil^2}{\mtil_0} =\frac{\mu_0^{2/3}\tilde\ell_0\cstil^2}{\mtil_0},
\eeq
from equation (\ref{eq:twothirds}). 
Including the
Heavy Ion Approximation 
($\tilde{\bar\chi_i}=\calr\bar\chi_i$ and $\tilde\gamma_{\rm AD}
\propto\gad/\calr$),
the ambipolar diffusion constant in the code is
\beq
\tilde\gamma_{\rm AD}=\frac{\gad\mcode\tcode}{\calr\lcode^3}
=\left[\frac{2\radlo}{\calr\bar\chi_i\calm\beta}\right]
\frac{\cstil\ltil_0^2}{\mtil_0},
\label{eq:gadtil}
\eeq
where the second step follows from equation (\ref{eq:radltwo}).
[Keep in mind that $\tilde\ell_0$, $\cstil$, and $\mtil_0$ are arbitrary; 
in the simulations described in the text, we have taken
$\tilde\ell_0=2$, $\cstil=0.1$ and $\tilde{\bar\rho}=\mtil_0/\tilde\ell^3_0=1$, 
so that $\tilde\gamma_{\rm AD}=
0.1\radlo/\calr\bar\chi_i\calm\beta$.] So long as the Heavy Ion Approximation is valid, 
the outcome of a simulation is independent of the value of $\calr$
since $\tilde\gamma_{\rm AD}$ always enters in combination with 
$\tilde{\bar\chi_i}=\calr\bar\chi_i$.

We can now address the issue of scaling in AD simulations
when the physical ionization is specified by, for example, the value of $\chiis$.
In carrying out a simulation, the ionization in code units,
$\tilde{\bar\chi_i}=\calr\bar\chi_i$, must be specified; hence, 
the physical ionization $\chiis\propto\bar\chi_i\propto \calr^{-1}$.
As noted above, the results of a simulation are independent of $\calr$ so
long as the Heavy Ion Approximation is valid. Thus, 
a single simulation provides the results for a family of problems with
different degrees of ionization 
but the same values
of $\radlo$ and $c_s$; as a result, we can use a single simulation 
to treat the physically plausible range of ionizations for a given value of
$\radlo$, as discussed in \S \ref{sec:adscale}.

\clearpage
\clearpage

\clearpage

\clearpage

\clearpage

\clearpage

\clearpage

\clearpage

\begin{table}
\begin{center}
\caption{AD Reynolds Number $\rad$ for Observed Molecular Clumps \citep{cru99} \label{tbl-1}}
\begin{tabular}{llccclcl}
\\
\tableline\tableline
Cloud        &$\beta$   &log $n_2$           &$R$  &$\calm$ &$\calm_A$ &$T_k$ &$\rad^a$ \\
             &          &(H$_2$ cm$^{-3}$) &(pc) &        &          &(K)   & \\
\tableline
W3 OH	     &0.07	  &6.8	&0.02	&1.9	&0.3 	&100	&3.0 \\
DR 21 OH1    &0.21	  &6.3	&0.05	&4	&1.3 	&50	&37.3 \\
Sgr B2	     &0.0008      &3.4	&22	&22	&0.4 	&70	&10.3 \\	
M17 SW	     &0.008       &4.5	&1	&7	&0.5 	&50	&6.3 \\	
W3 (main)    &0.13	  &5.5	&0.12	&4.8	&1.2 	&60	&24.1 \\	
S106	     &0.04	  &5.3	&0.07	&3.6	&0.5 	&30	&3.7  \\	
DR 21 OH2    &0.41	  &6	&0.05	&4	&1.8 	&50	&51.5 \\	
OMC-1	     &0.65	  &5.9	&0.05	&1.7	&1   	&100	&21.9  \\	
NGC 2024     &0.35	  &5	&0.2	&3.7	&1.6 	&25	&72.7 \\	
S88 B	     &0.056       &3.8	&0.7	&5.9	&1   	&40	&12.9 \\	
B1	     &0.17	  &4	&0.2	&3.6	&1.1 	&12	&15.7 \\	
W49 B	     &0.024       &3	&1	&5.9	&0.6 	&10	&6.3  \\	
W22	     &0.033       &3	&4	&3.5	&0.5 	&10	&20.5 \\	
W40	     &0.027       &2.7	&5	&10	&1.2 	&10	&42.4 \\	
$\rho$ Oph 1 &0.42        &3.2	&0.8	&3.5	&1.6 	&25	&41.6 \\	
OMCN-4	     &$>$0.47	  &6	&0.03	&2.9	&$>$1.4 &35	&$>$30.7 \\
Tau G	     &$>$0.042    &3	&1	&5.1	&$>$0.7 &10	&$>$9.5 \\	
L183	     &$>$0.052    &3.1	&0.3	&2.4	&$>$0.4	&10	&$>$1.9 \\	
L1647	     &$>$0.047    &3	&3	&9	&$>$1.4 &10	&$>$56.4 \\	
$\rho$ Oph 2 &$>$0.14     &3	&0.9	&3.2	&$>$0.8 &25	&$>$11.3 \\	
TMC-1	     &$>$0.063    &3	&1.9	&5.9	&$>$1   &10	&$>$31.4 \\	
L1495 W	     &$>$0.063    &3	&0.9	&3.9	&$>$0.7 &10	&$>$9.8 \\	
L134	     &$>$0.14	  &3.2	&0.3	&2.7	&$>$0.7 &10	&$>$6.3 \\	
TMC-1C	     &$>$1.3	  &4	&0.2	&2	&$>$1.6 &10	&$>$73.0 \\	
L1521	     &$>$0.13	  &3	&1.2	&3.9	&$>$1   &10	&$>$27.0 \\	
L889	     &$>$0.28	  &3	&2.4	&7.3	&$>$2.7 &13	&$>$191.1 \\	
Tau 16	     &$>$0.22	  &3	&1.2	&3.9	&$>$1.3 &10	&$>$45.7 \\	
\tableline
\end{tabular}
\end{center}
$^a$ $\rad$ computed using equation (\ref{eq:radladtext})
\end{table}
\clearpage

\begin{table}
\begin{center}
\caption{Model Parameters and Regimes of AD \label{tbl-2}}
\begin{tabular}{lccccc}
\\
\tableline\tableline
Model$^a$ & $\gad$ & $\rad(\ell_0)$ & $\rad(\ell_0)_t^b$ & Regime of AD \\
\tableline
m3c2r-1 &4        &0.12     &0.076	&III\\
m3c2r0  &40       &1.2      &0.70	&II $\sim$ III\\
m3c2r1  &400      &12       &10.1	&II\\
m3c2r2  &4000     &120      &103.2	&II\\
m3c2r3  &40000    &1200     &1022	&I\\
m3i  	&$\infty$ &$\infty$ &		&I\\
\tableline
\end{tabular}
\end{center}
$^a$ Models are labeled as ``mxcyrn," where $x$ is the thermal Mach number, $y=|\log\chio|$, and $n=\log(\radl/1.2)$. Model ``m3i" is an ideal MHD. Model m3c2r0 is the same as model m3c2h in LMKF.

$^b$ $\rad$ from models using time-dependent ionization (see \S 3).

$^c$ Root mean squared (rms) values.
\end{table}
\clearpage

\begin{table}
\begin{center}
\caption{Comparison of Clump Properties in Models with Different $\rad$ \label{tbl-3}}

\begin{tabular}{lccc}
\\
\tableline\tableline
Model	 & m3c2r-1 & m3c2r1 & m3c2r3 \\
\tableline
$\radlo$ & 0.12    & 12     & 1200 \\
$n_{vn}(k)^a$ & $1.96\pm0.02$  & $1.89\pm0.03$ & $1.48\pm0.05$ \\
$\Gamma_{\rm fit}^b$ & - & $-1.43\pm0.10$ & $-1.22\pm0.11$ \\
$\avg{\mu_{\Phi,c}} / \mu_{\Phi, 0}^c$ & $0.122\pm0.004$ & $0.120\pm0.005$ & $0.111\pm0.003$ \\
$\sigma({\avg{\mu_{\Phi,c}} / \mu_{\Phi, 0}})^d$ & $0.056$ & $0.074$ & $0.039$ \\
$\avg{\rho_c}$& $6.85\pm0.24$ & $5.42\pm0.23$ & $3.19\pm0.09$ \\
$\avg{\rcz/R_{c,\perp}}$ & $1.13\pm0.02$ & $1.15\pm0.02$ & $2.00\pm0.05$ \\
$\avg{r_c} (\rm cell)^e$ & $9.4\pm1.1$ & $12\pm1.5$ & $13.5\pm1.7$ \\
$\avg{N_c}$ ($d_c > 12$ cells)$^f$ & 698 & 434 & 349 \\
\tableline
\end{tabular}
\end{center}

$^a$ Velocity power spectral index of neutral component.

$^b$ Slope of the \clmf\ 
in the inertial range. 
The data for model m3c2r-1 do not have a single power law over the inertial range.

$^c$ Clump mass-to-flux ratio normalized by that of the whole box.

$^d$ Dispersion of clump mass-to-flux ratio normalized by that of the whole box.

$^e$ Mean radius of clumps in units of number of cells.

$^f$ Mean number of clumps with diameter larger than 12 cells.

\end{table}

\clearpage

\end{document}